\documentclass[10pt,twocolumn,letterpaper]{article}

\usepackage{cvpr}

\usepackage{graphicx}	
\usepackage{amsmath}	
\usepackage{amssymb}	
\usepackage{booktabs}
\usepackage{times}
\usepackage{microtype}
\usepackage{epsfig}
\usepackage{caption}
\usepackage{float}
\usepackage{placeins}
\usepackage{color, colortbl}
\usepackage{stfloats}
\usepackage{enumitem}
\usepackage{tabularx}
\usepackage{xstring}
\usepackage{multirow}
\usepackage{xspace}
\usepackage{url}
\usepackage{subcaption}
\usepackage{xcolor}
\usepackage[hang,flushmargin]{footmisc}

% ...

\newcommand{\R}[1]{{%
    \textbf{%
        \ifstrequal{#1}{1}{\textcolor{red}{R#1}}{%
        \ifstrequal{#1}{2}{\textcolor{blue}{R#1}}{%
        \ifstrequal{#1}{3}{\textcolor{magenta}{R#1}}{%
        \ifstrequal{#1}{4}{\textcolor{teal}{R#1}}{%
                           \textcolor{cyan}{R#1}%
        }}}}%
    }%
}}

\usepackage{xr-hyper}

\makeatletter
\newcommand*{\addFileDependency}[1]{
  \typeout{(#1)}
  \@addtofilelist{#1}
  \IfFileExists{#1}{}{\typeout{No file #1.}}
}

\makeatother

\definecolor{cvprblue}{rgb}{0.21,0.49,0.74}
\usepackage[pagebackref,breaklinks,colorlinks,allcolors=cvprblue]{hyperref}
\usepackage[capitalize]{cleveref}
\crefname{section}{Sec.}{Secs.}
\crefname{table}{Table}{Tables}
\crefname{figure}{Fig.}{Figs.}

\crefname{appendix}{Suppl.}{Suppls.} 

\usepackage{algorithm}
\usepackage{algpseudocode}
\newcolumntype{H}{>{\setbox0=\hbox\bgroup}c<{\egroup}@{}}

\usepackage{booktabs}
\usepackage{tabularx}
\usepackage{subcaption}
\newcommand{\ours}{OmniFlow}

\begin{document}
\def\paperID{0000}
\def\confName{CVPR}
\def\confYear{2025}

\def\paperTitle{Paper Title}

\def\authorBlock{
    Author 1\thanks{Equal contribution} \qquad
    Author 2\footnotemark[1] \qquad
    Author 3 \\
    Institute \\
    {\tt\small \{email, addresses\}@inst.edu}
}

\newif\ifreview \newcommand{\review}{\reviewtrue}
\newif\ifcamera \newcommand{\cameraready}{\cameratrue}
\newif\ifrebuttal \newcommand{\rebuttal}{\rebuttaltrue}

\title{\ours: Any-to-Any Generation with Multi-Modal Rectified Flows}
\def\paperID{4167} 

\author{
Shufan Li$^{*1}$, Konstantinos Kallidromitis$^{*2}$, Akash Gokul$^{*3}$, Zichun Liao${^1}$ \\ Yusuke Kato$^2$, Kazuki Kozuka $^2$, Aditya Grover$^1$  
\\ $^1$ UCLA~ $^2$Panasonic AI Research~ $^3$Salesforce AI Research
\\
{ \tt\small *Equal Contribution }
\\
{ \tt\small Correspondence to jacklishufan@cs.ucla.edu}
}


\twocolumn[
\vbox{
\maketitle
\centering
\includegraphics[width= 0.85\textwidth]{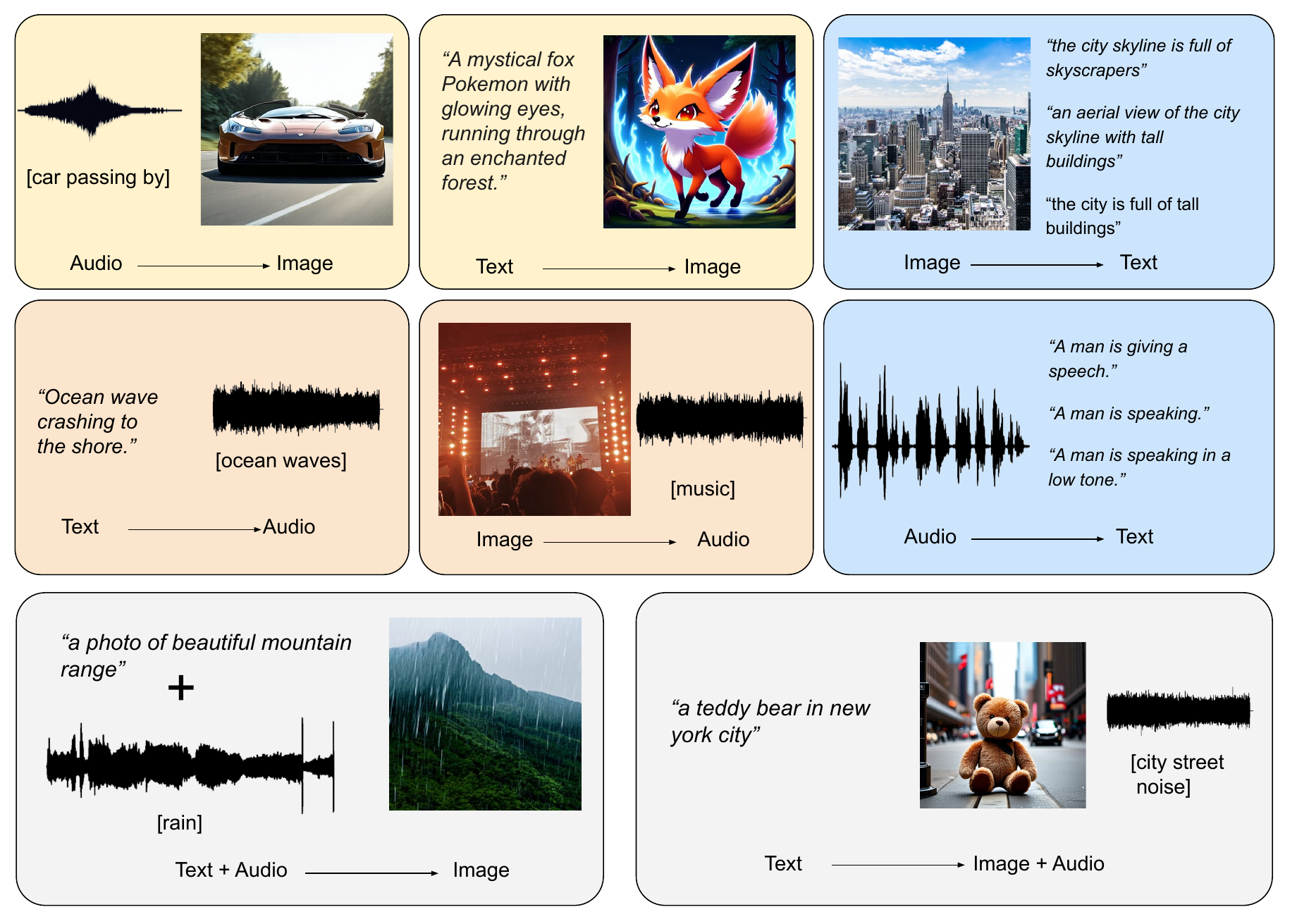}
\captionof{figure}{\textbf{\ours~is capable of a diverse range of any-to-any generation tasks}. \ours~supports generation of any output modalities given any input modality, such as text-to-image, text-to-audio, audio-to-image generations. It also supports tasks in multiple input modalities, such as text+audio-to-image. }
\label{fig:teaser}
\vspace{1em}}
]

\begin{abstract}

We introduce \ours, a novel generative model designed for any-to-any generation tasks such as text-to-image, text-to-audio, and audio-to-image synthesis. \ours~ advances the rectified flow (RF) framework used in text-to-image models to handle the joint distribution of multiple modalities. It outperforms previous any-to-any models on a wide range of tasks, such as text-to-image and text-to-audio synthesis. Our work offers three key contributions: First, we extend RF to a multi-modal setting and introduce a novel guidance mechanism, enabling users to flexibly control the alignment between different modalities in the generated outputs. Second, we propose a novel architecture that extends the text-to-image MMDiT architecture of Stable Diffusion 3 and enables audio and text generation. The extended modules can be efficiently pretrained individually and merged with the vanilla text-to-image MMDiT for fine-tuning. Lastly, we conduct a comprehensive study of the design choices of rectified flow transformers for large-scale audio and text generation, providing valuable insights into optimizing performance across various modalities.  Code is available at \href{https://github.com/jacklishufan/OmniFlows}{https://github.com/jacklishufan/OmniFlows}.

\end{abstract}

\section{Introduction}
Generative modeling has witnessed considerable advancements in recent years. Notably, diffusion models such as DALLE-3 \cite{openai2023dalle3}, Stable Diffusion 3 \cite{esser2024scaling}, AudioLDM2 \cite{liu2024audioldm} achieves state-of-the art performance on text-to-image and text-to-audio tasks. However, these models can only perform a single task while requiring considerable computing resources and data for training. To achieve any-to-any generations, previous works such as CoDi \cite{codi} and UIO \cite{lu2024unified} typically combine a set of modality-specific encoders (\eg ViT \cite{alexey2020image}) and decoders (\eg Stable Diffusion \cite{rombach2022high}). However, this design limits these models' ability to integrate information across modalities and generate multi-modal outputs coherently. For example, to perform audio+text-to-image (A+T$\rightarrow$I) generation, CoDi simply takes a weighted average of the audio embedding and text embedding to condition an image generator. However, there is no guarantee that the averaged embedding can faithfully represent the two input modalities, as arbitrarily many modality embeddings can average to the same embedding.

An alternative approach for any-to-any generation is to use a single multi-modal model to learn the joint distribution of multiple modalities. This approach has often led to strong performance as it allows information to flow across modalities. However, existing single-model designs typically involve training from scratch, and thus require a considerable amount of data. Existing works in this area, such as UniDiffuser \cite{unidiffuser} and Chameleon \cite{team2024chameleon} only experiment with text and image modalities. They also require considerable compute resources. To the best of our knowledge, there has yet to be a unified open-sourced multi-modal generative model that supports text, image, and audio simultaneously. 

We propose \ours, a unified multi-modal generative model for any-to-any generation. Unlike previous unified multi-modal models, \ours~ does not need to be trained from scratch with a large amount of data because of its modular design, saving considerable computing resources for its training. \ours~is inspired by the MMDiT architecture used in Stable Diffusion 3 \cite{esser2024scaling}, which performs text-to-image generation using a two-stream network that combines a text-input stream and an image-output stream through a series of joint attention blocks. \ours~builds on MMDIT by incorporating additional input and output streams,  extending its text-to-image capability to support any-to-any generation. Crucially, since the parameters for each stream are mostly independent, we can pretrain them separately or initialize them with a pretrained single-task expert model (\eg SD3). 

To effectively train \ours, we propose a novel multi-modal rectified flow formulation that incorporates a diverse set of tasks, such as text-to-audio and audio-to-image, into a unified learning objective. Multi-modal rectified flow is built upon a decoupled, time-differentiable interpretation between the distribution of a multi-modal data pair and i.i.d. Gaussian noise. In this formulation, each of the any-to-any generation tasks can be represented by a path connecting two noise levels. For example, given text, image, and audio modalities, the task of text+audio-to-image (T+A$\rightarrow$I) can be represented by a path between the distribution of (clean text, clean audio, Gaussian noise) to (clean text, clean audio, clean image).

We conducted extensive evaluations of \ours. Experiment results show that \ours~outperforms previous any-to-any models on a wide range of tasks, including text-to-image and text-to-audio generation. Compared to single-task specialist models, \ours~ achieves competitive performance with state-of-the-art methods. 

In summary, our contributions are three-fold:
\begin{itemize}
    \item First, we extend rectified flow formulation to the multi-modal setting and support flexible learning of any-to-any generation in a unified framework.
    \item Second, we proposed \ours, a novel modular multi-modal architecture for any-to-any generation tasks. It allows multiple modalities to directly interact with each other while being modular enough to allow individual components to be pretrained independently or initialized from task-specific expert models. 
    \item Lastly, to the best of our knowledge, we are the first work that provides a systematic investigation of the different ways of combining state-of-the-art flow-matching objectives with diffusion transformers for audio and text generation. We provide meaningful insights and hope to help the community develop future multi-modal diffusion models beyond text-to-image generation tasks.
\end{itemize}
\begin{figure}
    \centering
    \includegraphics[width=1.0\linewidth]{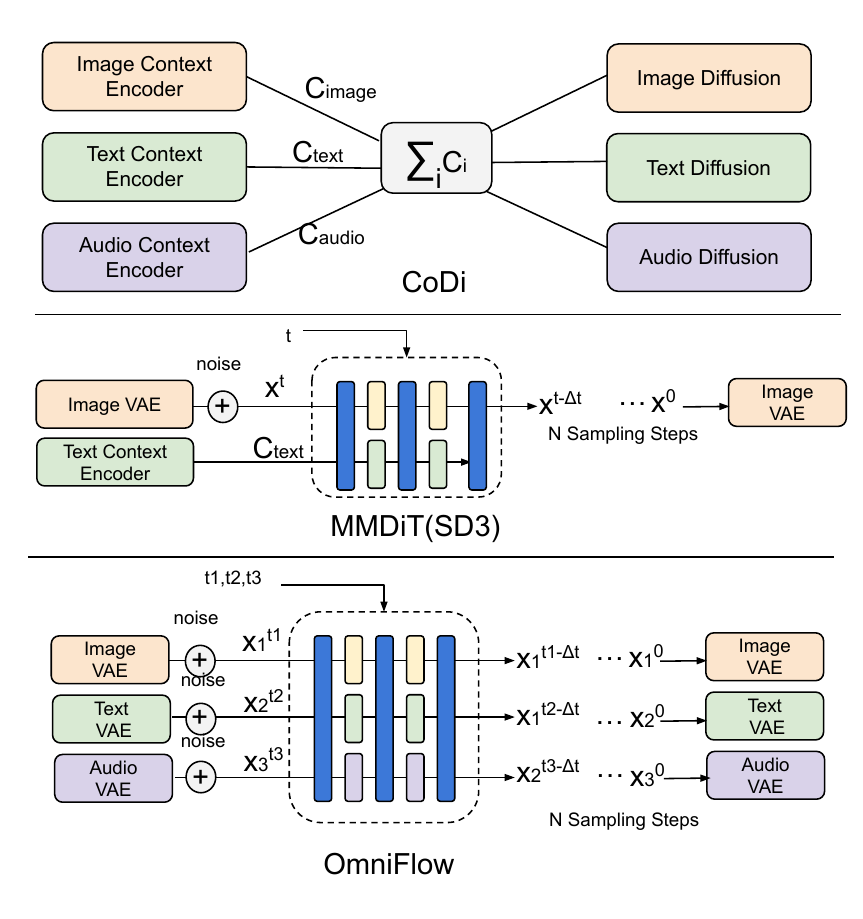}
    \caption{\textbf{Pipeline of \ours}. Previous any-to-any models such as CoDi \cite{codi} (Top) concatenate multiple modality-specific encoders and decoders, and naively average the embedding of multiple modalities to achieve joint conditioning. By contrast, \ours~(Bottom) is a unified, modular multi-modal model, where features from different modalities directly interact with each other through joint attention layers. \ours~is inspired by the modular design of Stable Diffusion 3 \cite{esser2024scaling} (Middle), a text-to-image model.}
    \label{fig:enter-label}
\end{figure}

\label{sec:intro}

\section{Backgrounds}
\label{sec:related}

\subsection{Flow-Based Generative Models}
\label{sec:background}
Flow-based generative models \cite{liu2022flow,lipman2022flow,klein2024equivariant,tong2023conditional}, represent the coupling of data points $x^0$ and noise distribution $x^1$ using an ordinary differential equation (ODE):
\begin{equation}
    dx^t=v_\theta(x^t,t)dt
\end{equation}

where the velocity $v$ is parameterized by a neural network. Directly solving this equation is expensive. However, we can define a forward process $x^t=a(t) x^0+b(t) x^1$ to directly regress a conditional vector field using the Conditional Flow Matching (CFM) objective \cite{tong2023conditional} as follows:

\begin{equation}
\mathcal{L_{\text{CFM}}}=(-\frac{b(t)\lambda'(t)}{2} )\mathbb{E}_{t,x^1,x^t|x^1}\lVert \epsilon_\theta(x^t,t)-x^1 \rVert^2
\end{equation}

where $\lambda(t)=\log\frac{\alpha(t)^2}{\beta(t)^2}$ is the signal-to-noise ratio (SNR),  $\epsilon_\theta(x^t,t)=-\frac{2}{\lambda'(t)b(t)}(v_\theta(x^t,t)-\frac{\alpha'(t)}{\alpha(t)}x^t)$ is parameterized by $v_\theta$. The optimum of this objective remains unchanged when introducing time-dependent weighting, and hence we can rewrite it following \cite{kingma2024understanding} as:

\begin{equation}
    L_w(x_0) = -\frac{1}2 \mathbb{E}_{t, \, x^1 } \,  w(t) \lambda'(t)  \|\epsilon_\Theta(z_t, t) - \epsilon \|^2
    \label{eq:unified-repr}
\end{equation}

where, $w(t)=-\frac{1}{2}\lambda'(t)b(t)^2$ for CFM and $x^1 \sim \mathcal{N}(0,I)$ follows noise distribution. This formulation gives a unified representation for a variety of generative modeling approaches. For example, a rectified flow's forward process is defined as $x^t=(1-t)x^0+tx^1$, which corresponds to $w^{\text{RF}}=\frac{t}{1-t}$. Esser et~al. \cite{esser2024scaling} summarized many configurations of common methods under this unified formulation, including (LDM)-Linear \cite{rombach2022high} and Cosine \cite{nichol2021improved}. They also explored a logit-normal distribution of timestep $t$ for text-to-image generation. We explore all these variants in the context of multi-modal generation, particularly for audio and text, as it is unclear if the results from the text-to-image domain can be directly generalized.

\subsection{Any-to-Any Generation}
Prior works have explored any-to-any generation. CoDi \cite{codi} achieved it first by combining multiple modality-specific encoders (\eg ViT) and decoders (\eg Stable Diffusion) through bridge alignment. However, its design has limited cross-modality interaction. For example, to achieve text+audio-to-image (T+A$\rightarrow$I generation), it simply computes the weighted average of text embeddings and audio embedding. Unified-IO \cite{lu2024unified} models any-to-any generation as a sequence-to-sequence problem, and uses an autoregressive model to achieve any-to-any generation, such as text-to-image or text-to-audio. Our work is the first to use a multi-modal flow matching objective for any-to-any tasks.

Additional works focus exclusively on unifying text-to-image and image-to-text generation. Chameleon \cite{team2024chameleon} uses an LLM-like large autoregressive model to handle multi-modal data. It represents images as VQGAN tokens \cite{yu2021vector}. Transfusion \cite{tranfusion} adopted a similar design, but uses a non-autoregressive diffusion loss for image modeling, while maintaining an autoregressive loss for text generation. Despite their successes, these unified multi-modal models require considerable training resources, because they are less modular than previous works that combine multiple models. \ours~achieves a good balance by separating the parameters of each individual modality, while allowing the features of each modality to freely interact with each other at every layer.

\section{Method}
\label{sec:method}

\subsection{Multi-Modal Rectified Flow}
\label{sec:multi-ref-method}

We consider the joint distribution $(x_1^0,x_2^0,..x_n^0) \sim \pi_{data}$ over the space of paired multi-modal data where $x_i\subseteq \mathbb{R}^{d_i}$ is a sample of modality $i$ represented by a vector of $d_i$ dimension. Let $(x_1^1,x_2^1,..x_n^1) \sim \pi^1$ be the i.i.d Gaussian distribution where $x_i^1 \sim \mathcal{N}(\textbf{0},\textbf{I})$ is a Gaussian vector of $d_i$ dimension. Given empirical observations $x^0 \sim \pi_{data}$, and $x^1 \sim \pi^1$, we consider the decoupled, continuous, time-differential interpolation given by:  

\begin{align}
\frac{\partial x_i^{t_i}}{\partial t_i} & = v_i(x_1^{t_1}, x_2^{t_2}, \dots, x_i^{t_i}, t_1, \dots, t_i) \\ \label{eq:multi-rf}
\frac{\partial x_i^{t_i}}{\partial t_j} & = 0 ; i \neq j \\
     x_i^{t_i} &=  (1-t_i) x^0_i + t_i x^1_i \label{eq:rf-fwd}
\end{align}

where the independence condition of Eq (2) indicates $x_i^{t_i}$ only moves when $t_i$ moves. Over this interpretation space, we can use a path $\tau:t\rightarrow(t_1,t_2..t_n);[0,1] \rightarrow [0,1]^n$ to model any-to-any generation tasks involving these modalities. For example, given $(x_1,x_2,x_3) \sim p_{data}$ where $x_1,x_2,x_3$ are image, text, and audio modalities. We can model text-to-image(T$\rightarrow$I) tasks as a path $\tau_{t2i}$ such that $\tau_{t2i}(0)=(0,0,1)$, which represents a clean text-image pair and $\tau_{t2i}(1)=(1,0,1)$, which represents clean text. We can similarly model the joint sampling of text, image and audio set as a path from $(0,0,0)$ to $(1,1,1)$ and text+image-to-audio ($T+I\rightarrow A$) as a path from $(0,0,0)$ to $(0,0,1)$.

The flow matching objective would be solving $n$ least squares regression problems for each modality of the form:

\begin{equation}
        \min_{v_\theta^i} \mathbb{E}_\tau  \int_\tau \mathbb{E}_{x^0,x^1} \lVert v_i - v_{\theta,i}(x_1^{t_1},x_2^{t_2},...x_n^{t_n},t_1..t_n) \rVert^2  ds
\end{equation}

where $v_i=x_i^0-x_i^1$, and $v_{\theta,i}$ is a neural network parameterized by $\theta$. We use the same network $\theta$ to predict outputs for all modalities $1,2..N$.  The outer expectation is over some prior of paths encoding generation tasks which we are interested in. The integral is calculated over a path $\tau(t)=(t_1,...t_n)$, and $ds=\frac{\partial t_i}{\partial t}dt$. Concretely, we consider three modalities: image, text, audio in our experiments as modalities: 1, 2 and 3 respectively. We consider the distribution of all possible linear paths $\tau(t)=(t_1,t_2,t_3)$ in $[0,1]^3$ following the rectified flow formulation. They can encode a diverse set of tasks such as text-to-image or text+image-to-audio. 

During training, we do not necessarily need all modalities for each data point. For data points that only contain a subset of three modalities (\eg text-image pairs), we can set the time step of remaining modalities (\eg audio) to 1, which corresponds to complete Gaussian noise. The full training algorithm is given as follows:

\begin{algorithm}[t]
\caption{Multi-Modal Rectified Flow}
\label{alg:syngen_step}
\begin{algorithmic}[1]
\Statex \textbf{Input:} Dataset $\mathcal{D}$ consists of modality $1,...N$,where each sample $x=(x_{i1}^0,x_{i2}^0,..)$ consists of a subset (or all) of modalities $i_1,i_2..\in\{1,2,..N\}$. 
\Statex \textbf{Output:} $v_{\theta,i}:(x_1^{t_1},x_2^{t_2},...x_n^{t_n})\rightarrow v_i^{t_i} $ for each $i=1,2..N$, parameterized by $\theta$
\Statex Initialize $\theta$
\While{not converged} 
    \State Sample $x=(x_{i1}^0,x_{i2}^0,..)\sim \mathcal{D}$
    \State $x_j^0 \gets \textbf{0};  \forall j\in\{1,2..N\}  \setminus \{i1,i2...\} $
     \State Sample path $\tau$.*
    \State Sample $t \sim \text{Uniform} ([0,1])$
    \State  $(t_1...t_N) \gets \tau (t) $
    \State $x_i^{t_i} \gets x_i^{t_i} =  (1-t_i) x^0_i + t_i x^1_i; \forall i \in {1,2..N} $
    \State $\mathcal{L}=\sum_{i\in\{i_1,i_2..\}} \lVert v_i - v_{\theta,i}(x_1^{t_1},...x_n^{t_n},t_1..t_n) \rVert^2$
    \State Perform optimizer step using $\nabla_\theta\mathcal{L}$
    \EndWhile
\State \textbf{Return} $\theta$
\State
\Comment{* $\tau$ encodes a task involving only modality $i_1,i_2..$, hence $t_j=1 ; \forall j \notin \{i_1,i_2..\}$}
\end{algorithmic}
\end{algorithm}

At inference, we simply pick a path and use the network prediction to solve for \cref{eq:multi-rf}. Notably, for standard text-to-image generation with $(x_1,x_2)$ pairs where $x_1$ is image and $x_2$ is text, and $x_3$ is the missing audio modality, picking a linear path from $(1,0,1)$ to $(0,0,1)$ is equivalent to the standard single-modality rectified flow (Text$\rightarrow$Image) formulation used by Stable Diffusion 3 \cite{esser2024scaling}.

\subsection{Multi-Modal Guidance}
\label{sec:guidance}

To flexibly control the multi-modal generation process, we extend the classifier free guidance (CFG)\cite{ho2022classifier} to multi-modal rectified flow setting. Recall that CFG of single modalities are formulated as follows:

\begin{equation}
    \hat{v}_\theta(x^t,c) =v_\theta(x^t,c) + (\alpha -1)(v_\theta(x^t,c)-v_\theta(x^t))
    \label{eq:single-cfg}
\end{equation}

where $c$ is a condition and $x^t$ is the noised latent at timestep $t$ of the single-modal output. We extend this formulation to multi-modal setting by defining $\delta_{ij}=v_\theta(x_i^t,x_j^0)-v_\theta(x_i^t)$, which represents the influence of input modality $j$ to output modality $i$. In particular, we obtain $v_\theta(x_i^t,x_j^0)$ and $v_\theta(x_i^t)$ by setting inputs of modalities not present in the formula to Gaussian noise. For example, given three modalities $x_1,x_2,x_3$, we can obtain $v_\theta(x_1^t,x_2^0)$  by computing $v_\theta(x_1^t,x_2^0,x_3^1)$ and  obtain $v_\theta(x_1^t)$ by computing $v_\theta(x_1^t,x_2^1,x_3^1)$. Note that $x_2^1,x_3^1$ is just Gaussian noise.

Given the set of $\delta_{ij}$, we can guide the output generation of modality $i$ by the following formula:

\begin{equation}
    \hat{v}_\theta(x^{t_1}_1...x^{t_n}_n) =v_\theta(x^{t_1}_1...x^{t_n}_n) + \sum_{j\neq i}(\alpha_{ij}-1)\delta_{ij}
    \label{eq:multi-rf-cfg}
\end{equation}

where $\alpha_{ij}$ is the equivalent of $\alpha$ in a multi-modal setting. This scheme allows the user to precisely control the interaction between each of the input and output modalities. When there are only two modalities, our multi-modal guidance \cref{eq:multi-rf-cfg} is equivalent to the standard single-modal classifier-free guidance \cref{eq:single-cfg}. 

\begin{figure*}[ht]
    \centering
    \begin{subfigure}[t]{0.45\textwidth}
        \centering
        \includegraphics[width=1.0\linewidth]{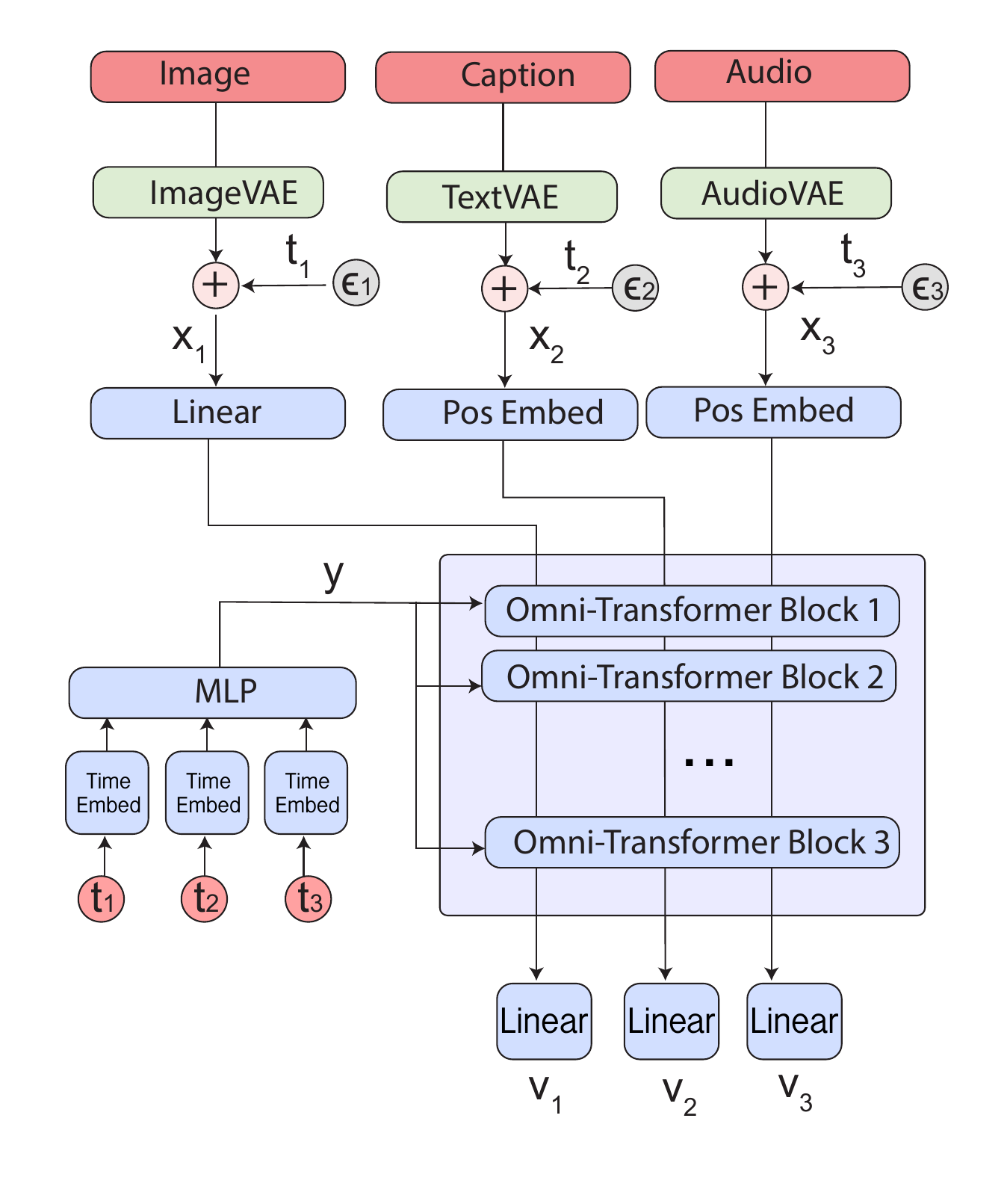}
        \caption{Overall Pipeline of \ours}
        \label{fig:subfig1}
    \end{subfigure}
    \begin{subfigure}[t]{0.45\textwidth}
        \centering
        \includegraphics[width=1.0\linewidth]{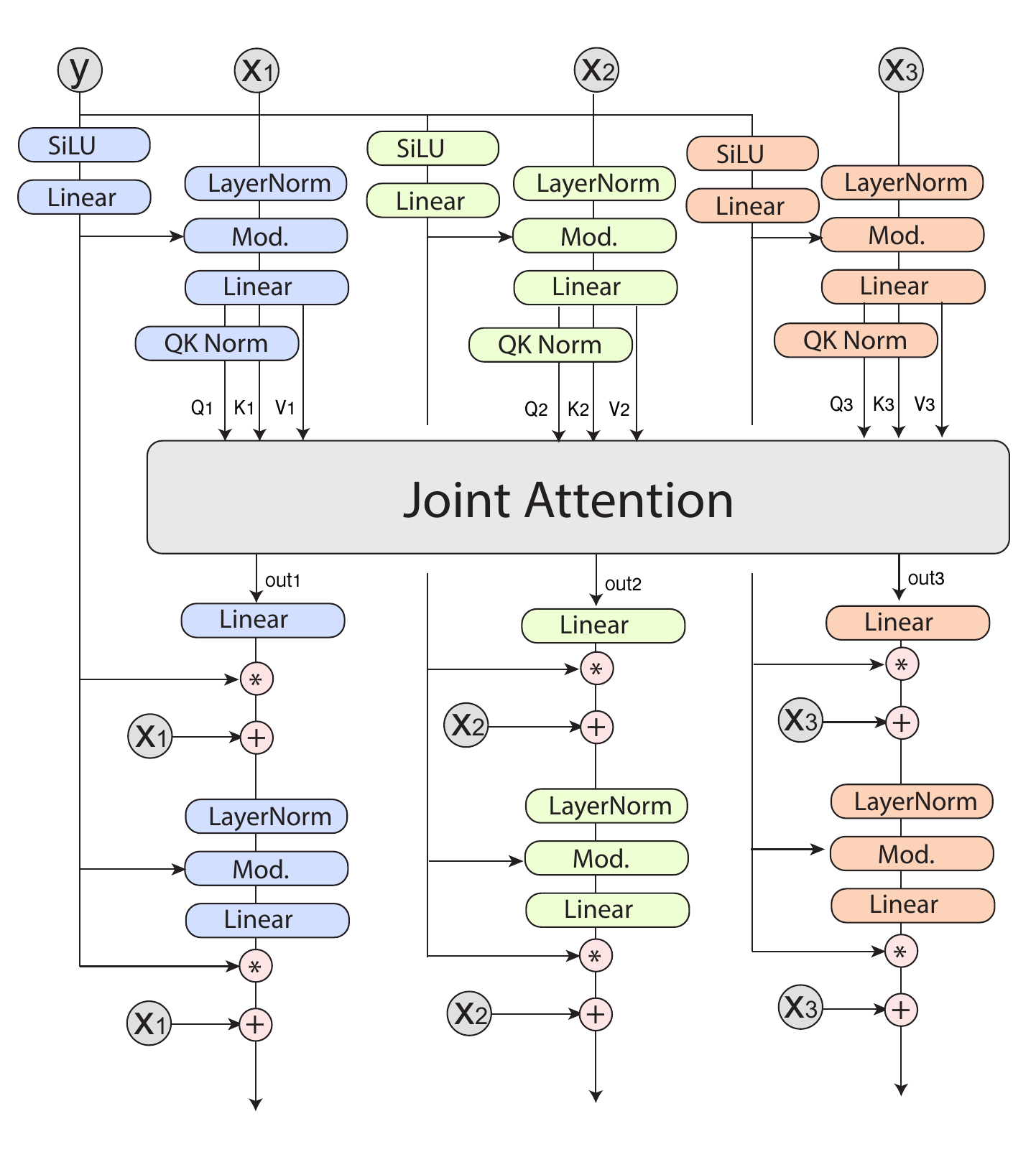}
        \caption{Design of Omni-Transformer Block}
        \label{fig:subfig2}
    \end{subfigure}

    \caption{\textbf{Architecture of \ours}. Left: We highlight the architecture of \ours.  Right: We show the design of an individual Omni-Transformer Block. }
     \label{fig:explainer-arch}
\end{figure*}
\subsection{Model Architecture}

We propose \ours, a modular, effective extension to the MMDiT architecture used in Stable Diffusion 3. Concretely, given multi-modal inputs that consist of text, image, and audio, we first convert them to latents $x_1,x_2,x_3$ using modality-specific VAEs. We then add random Gaussian noise to the latents following the forward process defined in \cref{eq:rf-fwd}. We use the three sinusoidal embeddings to encode, $t_1,t_2,t_3$ which correlate to the noise scale for each modality. These three timestep embeddings are passed to an MLP to obtain $y$, a single embedding representing all modality-specific time steps. The final input to \ours~are the unified timestep embedding y, and noised latents $(x_1,x_2,x_3)$. These four input vectors are passed to $N$ consecutive Omni-Transformer blocks. The final hidden states of each modality, are then processed by the linear output layer to obtain predictions of $v$.

Within each Omni-Transformer block, the inputs $x_1,x_2,x_3$ are processed by modality-specific projections to obtain $q_1,k_1,v_1,q_2,k_2,v_2,q_3,k_3,v_3$.  We then concatenate the queries, keys, and values to obtain $Q=\text{Concat}(q_1,q_2,q_3),K=\text{Concat}(k_1,k_2,k_3),V=\text{Concat}(v_1,v_2,v_3)$. The joint attention output for $i^{th}$ modality $\text{out}_i$  is given by:

\begin{equation}
    \text{out}_i=\text{SoftMax}(\frac{q_i^TK}{\sqrt{d}})V
\end{equation}

where $d$ is the dimension of each attention head. The output is passed to a feed forward network (FFN) to get the final output of the Omni-Transformer block. Following the design of DiT \cite{peebles2023scalable}, we use the unified time embedding to modulate the qkv projection and FFN. We add skip connections after the joint attention operation and after the FFN. 

We illustrate the model architecture in \cref{fig:explainer-arch}. Notably, different modalities are handled by different projection and feed-forward layers with independent parameters. The only multi-modal operation is the joint attention, with no trainable parameters of its own. This allows us to pretrain layers of different modalities individually and combine them for fine-tuning, which significantly improves the training efficiency.

\section{Setup}

\subsection{Training Dataset}

We use text-image pairs, text-audio pairs, and audio-image pairs during training. We also make use of a small amount of text-image-audio triplets.   The text-image pairs include 5M images sampled from COYO-700M dataset \cite{kakaobrain2022coyo-700m}, 2M images sampled from LAION-Aesthetic-3M subset \cite{laion_aesthetics}, 7M images from LAION-COCO subset \cite{laion_coco}, the full CC12M dataset \cite{changpinyo2021cc12m},  and 2M high-quality image dataset generated by flux-dev and DALLE-3 \cite{jackyhate_text_to_image_2m}. We put high weights on images from LAION-Aesthetic-3M and the 2M high-quality images to maintain good aesthetic quality in the output. The text-audio pairs include the full training set of AudioSet \cite{gemmeke2017audio}, Audiocaps \cite{kim2019audiocaps} and WavCaps \cite{mei2024wavcaps}. The audio-image pairs include the training data of VGGSound \cite{chen2020vggsound} and SoundNet \cite{aytar2016soundnet}. While SoundNet contains 2M images and is larger than VGGSound, we set the sample weight of VGGSound and SoundNet to 2:1 since SoundNet contains many improperly resized images with bad aspect ratios.

To generate text-image-audio triplets, we use BLIP \cite{li2022blip} to generate synthetic captions for videos in VGGSound and SoundNet. We provide further details of the dataset construction in the Appendix.

\subsection{Training Recipe}

\label{sec:training_receipe}

At a high level, we initialize \ours~with the text and image modules of Stable-Diffusion 3 (Model 1). We first train a separate text-to-audio model with text-audio pairs (Model 2). Then, we merge Model 1 and Model 2 to obtain a combined model with text, image, and audio modules (Model 3). Since Model 1 and Model 2 have separate text modules, we average their weights during the merge process. Finally, we fine-tune Model 3 on a diverse set of any-to-any tasks using the methods described in \cref{sec:multi-ref-method}. 

Due to our modular design, we can initialize and pretrain each module individually. This saves immense computational cost when compared to previous unified multi-modal models (\eg UniDiffuser \cite{unidiffuser}) which are trained from scratch. We use a global batch size of 64 and train Model 2 and Model 3 for 100k, and 150k steps each. We provide further training and implementation details in the Appendix.

\section{Main Results}

\subsection{Evaluation Metrics}
We perform extensive experiments on paired generation (text-to-image, text-to-audio) and generic any-to-any generation such as text-to-audio+image (T$\rightarrow$I+A), audio-to-text+image (A$\rightarrow$T+I). For text-to-image generation, we report FID \cite{heusel2017gans} and CLIP \cite{radford2021learning} scores on MSCOCO-30K benchmark \cite{lin2014microsoft}. Following the official implementation, the cosine similarities between CLIP embeddings are multiplied by 100. We also report results on the GenEval benchmark \cite{ghosh2024geneval}. For audio generation, we report FAD \cite{fad} and CLAP \cite{elizalde2023clap} score on AudioCaps. Results are reported with a 16kHz sampling rate. We also use CLAP scores for caption evaluations.

\subsection{Text-to-Image Generation}

\begin{table}[h!]
\centering
\begin{tabular}{l|c|c|c}

\textbf{Model} & \textbf{Param} & \textbf{FID$\downarrow$} & \textbf{CLIP$\uparrow$} \\
\hline

UniDiffuser & 0.9B & \textbf{9.71} & 30.93 \\
CoDi & 4.3B & 11.26 & 30.69 \\
\textcolor{gray}{UIO-2XXL} & \textcolor{gray}{6.8B} & \textcolor{gray}{13.39} & \textcolor{gray}{-} \\
\hline
SDv1.5 & 0.9B & 11.12 & 30.63 \\
SDXL* & 2.6B & 16.49 & 31.36 \\
SD3-Medium* & 2B & 20.94 & 30.65 \\
\hline
\ours* & 3.4B & 13.40 & \textbf{31.54} \\
\hline
\end{tabular}
\caption{\textbf{Text-to-Image Generation on MSCOCO-30K Benchmark.} *Indicates models pretraining data consists of high quality images and captions that do not follow the distribution of COCO dataset, which can negatively affect FID scores. }
\label{tab:fid}
\end{table}

\begin{table}[h!]
\centering
\begin{tabular}{cH|c|c|c}

\textbf{Model} & \textbf{M.} & \textbf{Param} & \textbf{Images}& \textbf{Gen.$\uparrow$} \\
\hline
\multicolumn{5}{c}{\textit{Text-to-Image Specialist}} \\
SD1.5 & I & 0.9B & 4.0B & .43 \\

SDv2.1 & I & 0.9B & 2.3B & .50 \\

SDXL & I & 2.6B & 1.6B & .55 \\

DALL-E 2 & I & 4.2B & 2.6B & .52 \\

SD3-Medium & I & 2B & 1B & .62 \\
\hline
\textcolor{gray}{SD3-Large} & \textcolor{gray}{I} & \textcolor{gray}{8B} & \textcolor{gray}{2.0B} & \textcolor{gray}{.68} \\
\hline
\multicolumn{5}{c}{\textit{Generalist}} \\

CoDi & I+T+A & 4.3B & 400M* &  .38 \\
UniDiff. & I+T & 0.9B & 2B & .43 \\
\ours & I+T+A & 3.4B & 30M* & \textbf{.62} \\
\hline

\textcolor{gray}{Chameleon} & I+T & \textcolor{gray}{7B} & \textcolor{gray}{3.5B} & \textcolor{gray}{.39} \\
\textcolor{gray}{Transfusion} & I+T & \textcolor{gray}{7B} & \textcolor{gray}{3.5B} & \textcolor{gray}{.63} \\
\hline
\end{tabular}
\caption{\textbf{Text-to-Image Generation on GenEval Benchmark.} We compare the model size, number of training images and GenEval benmark Score. * Indicates fine-tuning dataset. CoDi and MMDiT-O are both initialized with pretrained text-to-image diffusion models (SD and SD3). }
\label{tab:geneval}
\end{table}
We report results on MSCOCO-30k in \cref{tab:fid}, and results on GenEval in table \cref{tab:geneval}. On MSCOCO-30k, we achieve a lower FID than state-of-the-art models such as SDXL and SD3-Medium. While our FID number is higher than some previous models such as SDv1.5, it should be noted that more recent models such as SDXL and SD3 tend to have higher FID numbers because they are trained on high-quality text-image pairs that do not match the distribution of COCO images \cite{podell2023sdxl}. Notably, SD3 has a FID of 20.94 while SDv1.5 has 11.12, even though SD3 is considered a better model according to human evaluations. SDXL, which is widely recognized as the state-of-the-art open-source model before the release of SD3, also has a higher FID than SDv1.5.

In terms of CLIP scores, \ours~significantly outperforms previous models. In particular, when contrasted with generalist models UniDiffuser and CoDi, we achieve a gain of $+0.61$ and $+0.85$ respectively, showing superior text-to-image alignment. On GenEval Benchmark, which better measures the text-to-image capabilities, \ours~ achieves a score of 0.62, a competitive score even when compared to the state-of-the-art specialist SD3-Medium. In addition, \ours~significantly outperforms previous any-to-any baselines at the same scale, such as CoDi (+.24) and UniDiffuser (+.19). Compared with larger models trained on a lot more images, \ours~outperforms Chameleon-7B and achieves competitive performance as  Transfusion-7B.

Notably, unlike Chameleon, Transfusion, and UniDiffser which need to be trained from scratch, \ours~achieves high performance with only 30M training images, highlighting the effectiveness of our modular design. While the design of CoDi also allows it to make use of pretrained text-to-image model as its initialization, it is trained with considerably more images than \ours \ while performing worse.

\subsection{Text-to-Audio Generation}

\begin{table}[h!]
\centering
\begin{tabular}{l|c|c|l}

\textbf{Model} & \textbf{Param} & \textbf{FAD$\downarrow$} & \textbf{CLAP$\uparrow$} \\
\hline
\multicolumn{3}{c}{\textit{Text-to-Audio Specialist}} \\

AudioGen-L\cite{kreuk2022audiogen} &1B & 1.82 & - \\
Make-an-Audio\cite{huang2023make} &0.4B & 2.66 & - \\
AudioLDM-L\cite{liu2023audioldm} & 0.7B& 1.96 & .141 \\
Make-an-Audio 2\cite{huang2023make2}&0.9B & 2.05 & .173 \\
AudioLDM 2-Full-L\cite{liu2024audioldm}&0.7B & 1.86 & .182 \\

\hline
\multicolumn{3}{c}{\textit{Generalist}} \\
CoDi &3.4B& 1.80 & .053* \\
\ours &3.4B & \textbf{1.75} & \textbf{.183} \\
\textcolor{gray}{UIO-2XXL} &\textcolor{gray}{6.7B}& \textcolor{gray}{2.64} & \textcolor{gray}{-} \\
\hline
\end{tabular}
\caption{\textbf{Text-to-Audio Generation on AudioCaps Evaluation Set.} Comparison of FAD and CLAP scores for various audio generators. *Reproduced from official checkpoint, see Appendix for details.}
\label{tab:audio_models}
\end{table}

We report text-to-audio generation results on AudioCaps in \cref{tab:audio_models}.  Compared with previous state-of-the-art, \ours~achieves strong performance on FAD and CLAP scores. It outperforms AudioLDM2 on FAD (-0.11) and achieves equivalent performance on CLAP (+0.001). When compared with generalist models, \ours~significantly outperforms CoDi on both FAD (-0.05) and CLAP (+.13) metrics.

\subsection{Recipes for Audio and Text Diffusions}

\begin{table}[h!]
\centering
\begin{tabular}{l|c|c}

\textbf{} & \textbf{Audio Gen.}  &  \textbf{Text Gen.} \\
\textbf{} & \textbf{FAD$\downarrow$}  &  \textbf{CLAP$\uparrow$} \\
\hline
\multicolumn{3}{c}{\textit{Continuous Flow Matching}} \\
eps/linear & 2.08 & .141\\
v/cos  & 2.01 & .203 \\
v/linear & 1.86 & .126 \\
\hline
rf/uniform & 1.82  & .227 \\
rf/lognorm & \textbf{1.79} & \textbf{.254}\\
\hline
\multicolumn{3}{c}{\textit{Discrete Text Diffusion}} \\
SEDD\cite{loudiscrete} & - & .180 \\
MDLM\cite{sahoo2024simple} & - &  .163 \\
\end{tabular}
\caption{\textbf{Various Formulations for Audio and Text Generation.} We report FAD for audio generation and CLAP for text generation on AudioCaps dataset. }
\label{tab:ablation}
\end{table}

We explore various recipes for training audio and text diffusion transformers for multi-modal generation, which is a relatively under-explored area. Concretely, we explored five formulations mentioned in the section \cref{sec:background}.  For these experiments, we used a model with only audio and text modules (Model 2 in \cref{sec:training_receipe}) and trained for 50k steps.  We report FAD score for text-to-audio generation and CLAP score for audio-to-text generation. Amongst all five formulations, rf/lognorm performs the best with the lowest FAD (1.79) and highest CLAP score (.254). We also explored two discrete space diffusion models, SEDD \cite{loudiscrete} and MDLM \cite{sahoo2024simple} which showed advantages over continuous-space diffusion models in recent literature. Specifically, we use the absorbing state version of SEDD. For these experiments, the text-vae encoder is replaced with a token-embedding layer, and, text-vae decoder is replaced with a simple linear output layer to predict token logits. We also replace the flow-matching loss on the text-embedding with the loss function of SEDD and MDLM respectively, which operates on token logits instead of continuous embeddings. We report the CLAP score on audio-to-text generation. We do not see considerable advantages over continuous alternatives. 

\section{Sampling}
On the sampling side, we explored the effect of guidance and timestep shift. The timestep shift was originally introduced by SD3 to balance the sampling process of images at different resolutions. Concretely, it augments the inference schedule as:
\begin{equation}
    \hat{t}=\frac{\gamma t}{1+(1-\gamma) t}
\end{equation}

where $\gamma=\sqrt{\frac{m}{n}}$, with $m$ being the target sample resolution and $n$ being a reference resolution. For audio and text generation, there is no concept of varying resolution, as the input audio spectrogram and text embedding have fixed resolutions. However, we empirically observe applying a shift can improve the generation quality. Concretely, incorporating the shift term $\gamma>1$ will lead to a concave schedule, where the denoising process progresses slowly at the beginning and accelerates towards the end. We find that this improves sample quality for text-to-audio and audio-to-text generation tasks.

\begin{figure}[ht]
    \centering
    \begin{subfigure}{0.23\textwidth}
        \centering
        \includegraphics[width=1.0\linewidth]{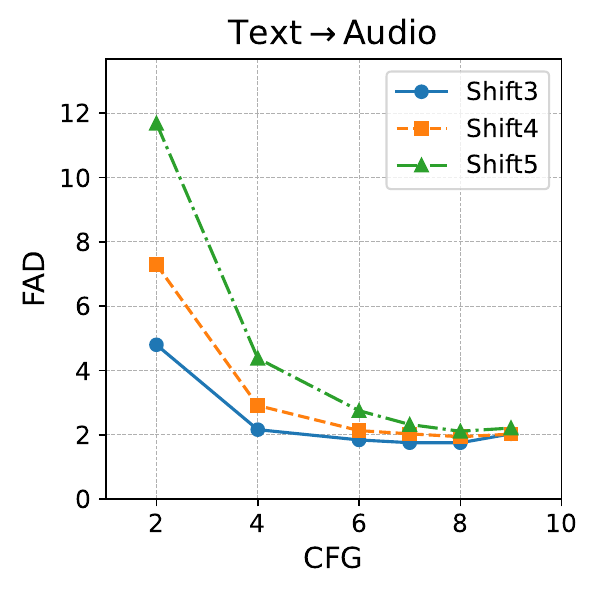}
        \caption{Text-to-Audio Generation.}
        \label{fig:cfg-subfig1}
    \end{subfigure}
    \begin{subfigure}{0.23\textwidth}
        \centering
        \includegraphics[width=1.0\linewidth]{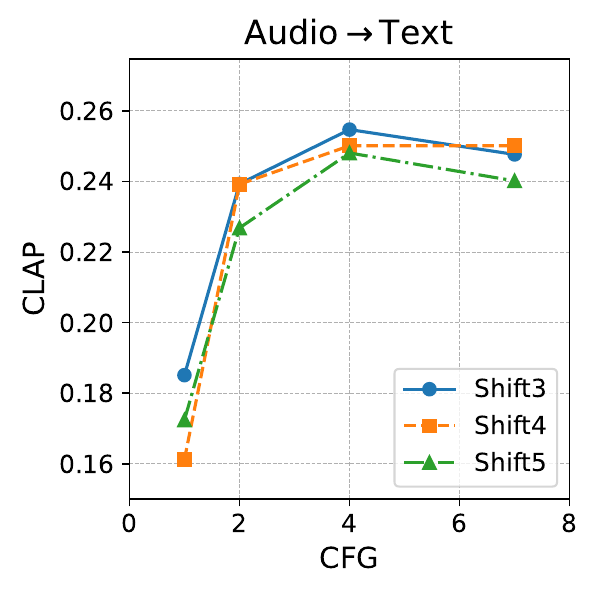}
        \caption{Audio-to-Text Generation.}
        \label{fig:cfg-subfig2}
    \end{subfigure}

    \caption{\textbf{Effect of CFG and Shift for audio and text generation}. We evaluate the impact of guidance and timestep shift on text-to-audio and audio-to-text tasks.}
    \label{fig:cfg}
\end{figure}

We employ the multi-modal guidance mentioned in \cref{sec:guidance}. For simple audio-to-text and text-to-audio generation, our formulation is reduced to standard classifier-free guidance. We show the effect of guidance and timestep shift in \cref{fig:cfg}. Generally, we find that shift=3.0 works well for both tasks. For audio generation, a guidance scale of 8 achieves the highest performance. For text generation, a guidance scale of 4 achieves the best result.  

\begin{figure}[t]
    \centering
    \includegraphics[width=1.0\linewidth]{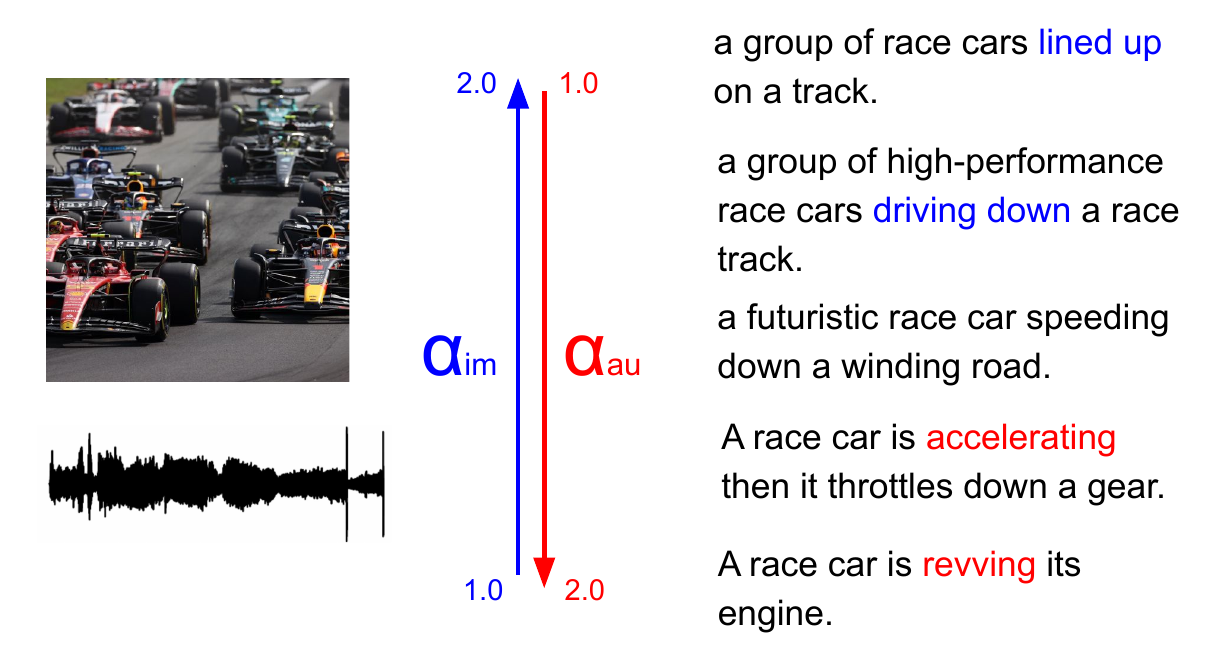}
    \caption{\textbf{Effect of Multi-Modal Guidance.} In this example, the user can flexibly control the alignment between output text and input image, audio independently by varying $\alpha_{\text{au}}$ and $\alpha_{\text{im}}$. Higher $\alpha_{\text{im}}$ will make the output texts resemble image captions, with visual descriptions such as \textcolor{blue}{lined up}, \textcolor{blue}{driving down}. Higher $\alpha_{\text{au}}$ will make the output texts resemble audio captions, with descriptions such as \textcolor{red}{accelerating}, \textcolor{red}{revving}.}
    \label{fig:multi-modal-cfg}
\end{figure}

To explore the effect of multi-modal guidance in \cref{sec:guidance}, we provide qualitative results for audio+image-to-text (A+I$\rightarrow$T) task. Recall that we use $x_1,x_2,x_3$ to denote image, text, and audio modalities. The multi-modal guidance for this task can be controlled by $\alpha_{21}$ and $\alpha_{23}$ where $\alpha_{21}$ controls text-image alignment and $\alpha_{23}$ controls text-audio alignment.  For simplicity, we denote $\alpha_{21}$ as $\alpha_{\text{im}}$ and $\alpha_{23}$ as $\alpha_{\text{au}}$. We vary $\alpha_{\text{im}}$, $\alpha_{\text{au}}$ between the interval $[1.0,2.0]$ such that $\alpha_{\text{im}}+\alpha_{\text{au}}=3.0$. We show the results in \cref{fig:multi-modal-cfg}. Qualitatively, higher $\alpha_{\text{au}}$ will make the model's output resemble more the audio captions, and $\alpha_{\text{im}}$ will make the model's output resembles more the image captions. Interestingly, we observe that it also reflects the subtle differences in the style of audio and image captions in the training data (\eg whether the first letter is capitalized). By varying these two parameters, users can achieve flexible control of generation.

\begin{figure}[h]
    \centering
    \includegraphics[width=1.0\linewidth]{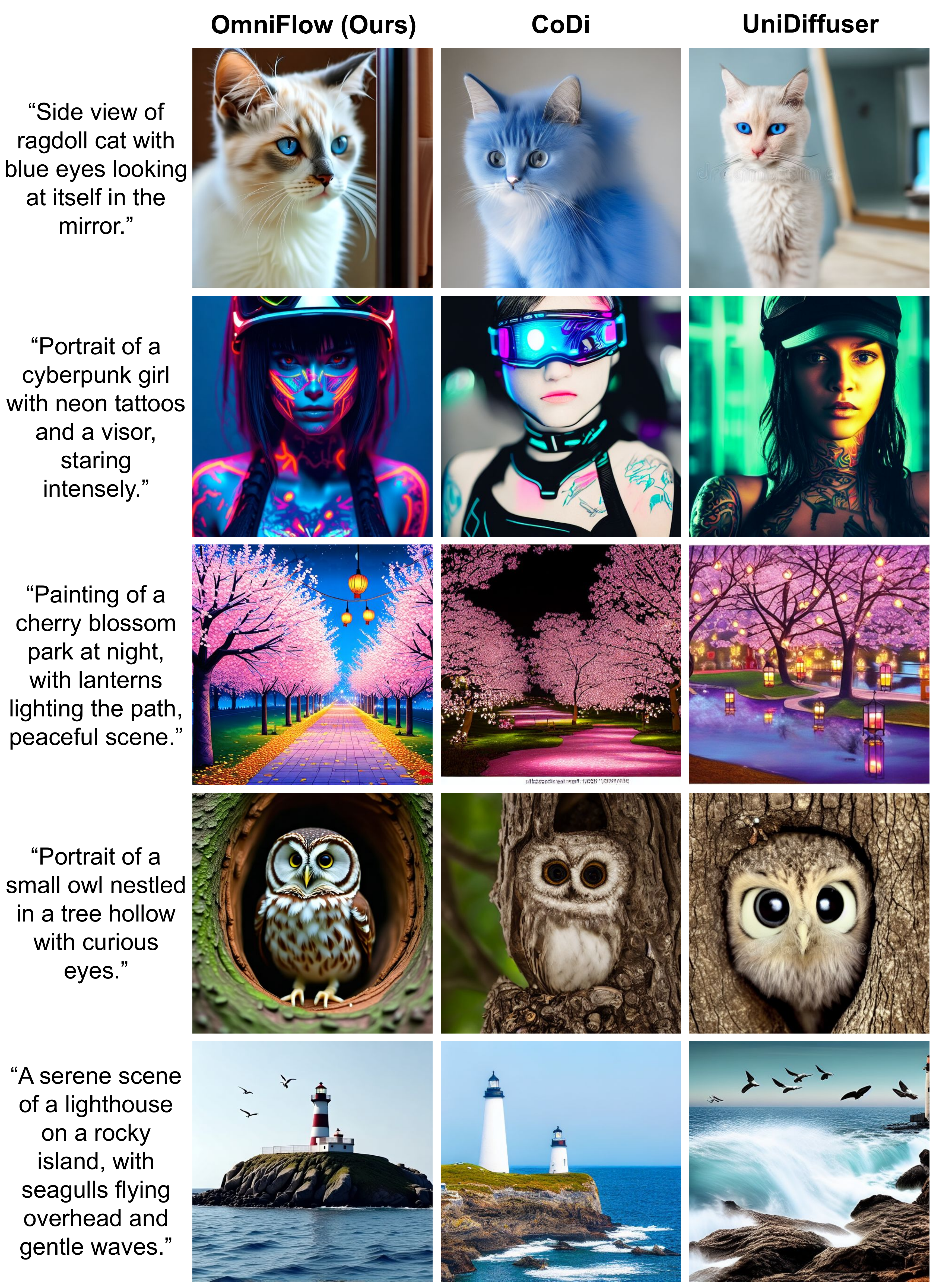}
    \caption{\textbf{Qualitative Comparison with baselines on text-to-image generation.} \ours~achieves better image  quality and prompt alignment when compared to previous generalist models.}
    \label{fig:side-by-side-t2i}
\end{figure}

\subsection{Qualitative Comparison}

We directly compare \ours~with two recent any-to-any generation methods: CoDi \cite{codi} and UniDiffuser \cite{unidiffuser}. In addition to the quantitative results, we present qualitative text-to-image comparisons in \cref{fig:side-by-side-t2i}. These examples demonstrate that \ours~achieves a significant improvement in generation quality compared to previous any-to-any models. Specifically, in the first example (top), our model successfully follows the prompt while maintaining high aesthetic quality, accurately capturing both the cat’s features and its mirrored reflection. In contrast, CoDi is unable to change the cat’s eyes, and UniDiffuser fails to depict the cat looking at the mirror. A similar trend is evident in the third example: \ours~correctly positions lanterns tied to a rope, while UniDiffuser places them on the river. Finally, in the lighthouse example, CoDi fails to incorporate seagulls, and UniDiffuser ignores the adjective “gentle,” instead producing an image with rough waves and an out-of-focus lighthouse.

Our results show that \ours~achieves a much higher generation quality compared with previous any-to-any models, both in terms of image-text alignment and image fidelity.

\section{Conclusion}
\label{sec:conclusion}

We present \ours, a unified early-fusion multi-modal generative model for any-to-any generation tasks. \ours~adapts a modular design that enables individual components to be pretrained separately, while allowing features from different modalities to directly interact with each other, through a joint attention mechanism. We conduct extensive experiments to show that \ours~ outperforms previous any-to-any models on a wide range of challenging generation tasks, including text-to-image and text-to-audio generation. We provide further analysis on the limitation of \ours~in the Appendix.  

\section{Acknowledgments}
This research was supported by NSF Career \#2341040 and a Schmidt Science Fellowship.

{\small
\bibliographystyle{ieeenat_fullname}
\bibliography{11_references}

\begin{thebibliography}{53}
\providecommand{\natexlab}[1]{#1}
\providecommand{\url}[1]{\texttt{#1}}
\expandafter\ifx\csname urlstyle\endcsname\relax
  \providecommand{\doi}[1]{doi: #1}\else
  \providecommand{\doi}{doi: \begingroup \urlstyle{rm}\Url}\fi

\bibitem[Alexey(2020)]{alexey2020image}
Dosovitskiy Alexey.
\newblock An image is worth 16x16 words: Transformers for image recognition at scale.
\newblock \emph{arXiv preprint arXiv: 2010.11929}, 2020.

\bibitem[Aytar et~al.(2016)Aytar, Vondrick, and Torralba]{aytar2016soundnet}
Yusuf Aytar, Carl Vondrick, and Antonio Torralba.
\newblock Soundnet: Learning sound representations from unlabeled video.
\newblock \emph{Advances in neural information processing systems}, 29, 2016.

\bibitem[BAI et~al.(2024)BAI, Liu, Wang, Shi, Wang, Plumbley, Gan, and Chen]{bai2024audiosetcaps}
JISHENG BAI, Haohe Liu, Mou Wang, Dongyuan Shi, Wenwu Wang, Mark~D Plumbley, Woon-Seng Gan, and Jianfeng Chen.
\newblock Audiosetcaps: Enriched audio captioning dataset generation using large audio language models.
\newblock In \emph{Audio Imagination: NeurIPS 2024 Workshop AI-Driven Speech, Music, and Sound Generation}, 2024.

\bibitem[Bao et~al.(2023)Bao, Nie, Xue, Li, Pu, Wang, Yue, Cao, Su, and Zhu]{unidiffuser}
Fan Bao, Shen Nie, Kaiwen Xue, Chongxuan Li, Shi Pu, Yaole Wang, Gang Yue, Yue Cao, Hang Su, and Jun Zhu.
\newblock One transformer fits all distributions in multi-modal diffusion at scale.
\newblock In \emph{International Conference on Machine Learning}, pages 1692--1717. PMLR, 2023.

\bibitem[Byeon et~al.(2022)Byeon, Park, Kim, Lee, Baek, and Kim]{kakaobrain2022coyo-700m}
Minwoo Byeon, Beomhee Park, Haecheon Kim, Sungjun Lee, Woonhyuk Baek, and Saehoon Kim.
\newblock Coyo-700m: Image-text pair dataset.
\newblock \url{https://github.com/kakaobrain/coyo-dataset}, 2022.

\bibitem[Changpinyo et~al.(2021)Changpinyo, Sharma, Ding, and Soricut]{changpinyo2021cc12m}
Soravit Changpinyo, Piyush Sharma, Nan Ding, and Radu Soricut.
\newblock {Conceptual 12M}: Pushing web-scale image-text pre-training to recognize long-tail visual concepts.
\newblock In \emph{CVPR}, 2021.

\bibitem[Chen et~al.(2020)Chen, Xie, Vedaldi, and Zisserman]{chen2020vggsound}
Honglie Chen, Weidi Xie, Andrea Vedaldi, and Andrew Zisserman.
\newblock Vggsound: A large-scale audio-visual dataset.
\newblock In \emph{ICASSP 2020-2020 IEEE International Conference on Acoustics, Speech and Signal Processing (ICASSP)}, pages 721--725. IEEE, 2020.

\bibitem[Chen et~al.(2024)Chen, Ma, Li, Xu, Liang, Zheng, Yu, and Chen]{chen2024slam}
Wenxi Chen, Ziyang Ma, Xiquan Li, Xuenan Xu, Yuzhe Liang, Zhisheng Zheng, Kai Yu, and Xie Chen.
\newblock Slam-aac: Enhancing audio captioning with paraphrasing augmentation and clap-refine through llms.
\newblock \emph{arXiv preprint arXiv:2410.09503}, 2024.

\bibitem[Chung et~al.(2024)Chung, Hou, Longpre, Zoph, Tay, Fedus, Li, Wang, Dehghani, Brahma, et~al.]{chung2024scaling}
Hyung~Won Chung, Le Hou, Shayne Longpre, Barret Zoph, Yi Tay, William Fedus, Yunxuan Li, Xuezhi Wang, Mostafa Dehghani, Siddhartha Brahma, et~al.
\newblock Scaling instruction-finetuned language models.
\newblock \emph{Journal of Machine Learning Research}, 25\penalty0 (70):\penalty0 1--53, 2024.

\bibitem[Elizalde et~al.(2023)Elizalde, Deshmukh, Al~Ismail, and Wang]{elizalde2023clap}
Benjamin Elizalde, Soham Deshmukh, Mahmoud Al~Ismail, and Huaming Wang.
\newblock Clap learning audio concepts from natural language supervision.
\newblock In \emph{ICASSP 2023-2023 IEEE International Conference on Acoustics, Speech and Signal Processing (ICASSP)}, pages 1--5. IEEE, 2023.

\bibitem[Esser et~al.(2024)Esser, Kulal, Blattmann, Entezari, M{\"u}ller, Saini, Levi, Lorenz, Sauer, Boesel, et~al.]{esser2024scaling}
Patrick Esser, Sumith Kulal, Andreas Blattmann, Rahim Entezari, Jonas M{\"u}ller, Harry Saini, Yam Levi, Dominik Lorenz, Axel Sauer, Frederic Boesel, et~al.
\newblock Scaling rectified flow transformers for high-resolution image synthesis.
\newblock In \emph{Forty-first International Conference on Machine Learning}, 2024.

\bibitem[Gemmeke et~al.(2017)Gemmeke, Ellis, Freedman, Jansen, Lawrence, Moore, Plakal, and Ritter]{gemmeke2017audio}
Jort~F Gemmeke, Daniel~PW Ellis, Dylan Freedman, Aren Jansen, Wade Lawrence, R~Channing Moore, Manoj Plakal, and Marvin Ritter.
\newblock Audio set: An ontology and human-labeled dataset for audio events.
\newblock In \emph{2017 IEEE international conference on acoustics, speech and signal processing (ICASSP)}, pages 776--780. IEEE, 2017.

\bibitem[Ghosh et~al.(2024)Ghosh, Hajishirzi, and Schmidt]{ghosh2024geneval}
Dhruba Ghosh, Hannaneh Hajishirzi, and Ludwig Schmidt.
\newblock Geneval: An object-focused framework for evaluating text-to-image alignment.
\newblock \emph{Advances in Neural Information Processing Systems}, 36, 2024.

\bibitem[Hate(2024)]{jackyhate_text_to_image_2m}
Jacky Hate.
\newblock Text-to-image-2m dataset, 2024.
\newblock Accessed: 2024-11-14.

\bibitem[Heusel et~al.(2017)Heusel, Ramsauer, Unterthiner, Nessler, and Hochreiter]{heusel2017gans}
Martin Heusel, Hubert Ramsauer, Thomas Unterthiner, Bernhard Nessler, and Sepp Hochreiter.
\newblock Gans trained by a two time-scale update rule converge to a local nash equilibrium.
\newblock \emph{Advances in neural information processing systems}, 30, 2017.

\bibitem[Ho and Salimans(2022)]{ho2022classifier}
Jonathan Ho and Tim Salimans.
\newblock Classifier-free diffusion guidance.
\newblock \emph{arXiv preprint arXiv:2207.12598}, 2022.

\bibitem[Ho et~al.(2022)Ho, Chan, Saharia, Whang, Gao, Gritsenko, Kingma, Poole, Norouzi, Fleet, et~al.]{ho2022imagen}
Jonathan Ho, William Chan, Chitwan Saharia, Jay Whang, Ruiqi Gao, Alexey Gritsenko, Diederik~P Kingma, Ben Poole, Mohammad Norouzi, David~J Fleet, et~al.
\newblock Imagen video: High definition video generation with diffusion models.
\newblock \emph{arXiv preprint arXiv:2210.02303}, 2022.

\bibitem[Huang et~al.(2023{\natexlab{a}})Huang, Ren, Huang, Yang, Ye, Zhang, Liu, Yin, Ma, and Zhao]{huang2023make2}
Jiawei Huang, Yi Ren, Rongjie Huang, Dongchao Yang, Zhenhui Ye, Chen Zhang, Jinglin Liu, Xiang Yin, Zejun Ma, and Zhou Zhao.
\newblock Make-an-audio 2: Temporal-enhanced text-to-audio generation.
\newblock \emph{arXiv preprint arXiv:2305.18474}, 2023{\natexlab{a}}.

\bibitem[Huang et~al.(2023{\natexlab{b}})Huang, Huang, Yang, Ren, Liu, Li, Ye, Liu, Yin, and Zhao]{huang2023make}
Rongjie Huang, Jiawei Huang, Dongchao Yang, Yi Ren, Luping Liu, Mingze Li, Zhenhui Ye, Jinglin Liu, Xiang Yin, and Zhou Zhao.
\newblock Make-an-audio: Text-to-audio generation with prompt-enhanced diffusion models.
\newblock In \emph{International Conference on Machine Learning}, pages 13916--13932. PMLR, 2023{\natexlab{b}}.

\bibitem[Kilgour et~al.(2018)Kilgour, Zuluaga, Roblek, and Sharifi]{fad}
Kevin Kilgour, Mauricio Zuluaga, Dominik Roblek, and Matthew Sharifi.
\newblock Fr$\backslash$'echet audio distance: A metric for evaluating music enhancement algorithms.
\newblock \emph{arXiv preprint arXiv:1812.08466}, 2018.

\bibitem[Kim et~al.(2019)Kim, Kim, Lee, and Kim]{kim2019audiocaps}
Chris~Dongjoo Kim, Byeongchang Kim, Hyunmin Lee, and Gunhee Kim.
\newblock Audiocaps: Generating captions for audios in the wild.
\newblock In \emph{Proceedings of the 2019 Conference of the North American Chapter of the Association for Computational Linguistics: Human Language Technologies, Volume 1 (Long and Short Papers)}, pages 119--132, 2019.

\bibitem[Kingma and Gao(2024)]{kingma2024understanding}
Diederik Kingma and Ruiqi Gao.
\newblock Understanding diffusion objectives as the elbo with simple data augmentation.
\newblock \emph{Advances in Neural Information Processing Systems}, 36, 2024.

\bibitem[Klein et~al.(2024)Klein, Kr{\"a}mer, and No{\'e}]{klein2024equivariant}
Leon Klein, Andreas Kr{\"a}mer, and Frank No{\'e}.
\newblock Equivariant flow matching.
\newblock \emph{Advances in Neural Information Processing Systems}, 36, 2024.

\bibitem[Kreuk et~al.(2022)Kreuk, Synnaeve, Polyak, Singer, D{\'e}fossez, Copet, Parikh, Taigman, and Adi]{kreuk2022audiogen}
Felix Kreuk, Gabriel Synnaeve, Adam Polyak, Uriel Singer, Alexandre D{\'e}fossez, Jade Copet, Devi Parikh, Yaniv Taigman, and Yossi Adi.
\newblock Audiogen: Textually guided audio generation.
\newblock \emph{arXiv preprint arXiv:2209.15352}, 2022.

\bibitem[LAION(2023{\natexlab{a}})]{laion_aesthetics}
LAION.
\newblock Aesthetics for open source, 2023{\natexlab{a}}.
\newblock Accessed: 2024-11-14.

\bibitem[LAION(2023{\natexlab{b}})]{laion_coco}
LAION.
\newblock Laion coco: 600m synthetic captions from laion2b-en, 2023{\natexlab{b}}.
\newblock Accessed: 2024-11-14.

\bibitem[Li et~al.(2020)Li, Gao, Li, Peng, Li, Zhang, and Gao]{li2020optimus}
Chunyuan Li, Xiang Gao, Yuan Li, Baolin Peng, Xiujun Li, Yizhe Zhang, and Jianfeng Gao.
\newblock Optimus: Organizing sentences via pre-trained modeling of a latent space.
\newblock \emph{arXiv preprint arXiv:2004.04092}, 2020.

\bibitem[Li et~al.(2022)Li, Li, Xiong, and Hoi]{li2022blip}
Junnan Li, Dongxu Li, Caiming Xiong, and Steven Hoi.
\newblock Blip: Bootstrapping language-image pre-training for unified vision-language understanding and generation.
\newblock In \emph{International conference on machine learning}, pages 12888--12900. PMLR, 2022.

\bibitem[Li et~al.(2023)Li, Li, Savarese, and Hoi]{li2023blip}
Junnan Li, Dongxu Li, Silvio Savarese, and Steven Hoi.
\newblock Blip-2: Bootstrapping language-image pre-training with frozen image encoders and large language models.
\newblock In \emph{International conference on machine learning}, pages 19730--19742. PMLR, 2023.

\bibitem[Lin et~al.(2014)Lin, Maire, Belongie, Hays, Perona, Ramanan, Doll{\'a}r, and Zitnick]{lin2014microsoft}
Tsung-Yi Lin, Michael Maire, Serge Belongie, James Hays, Pietro Perona, Deva Ramanan, Piotr Doll{\'a}r, and C~Lawrence Zitnick.
\newblock Microsoft coco: Common objects in context.
\newblock In \emph{Computer Vision--ECCV 2014: 13th European Conference, Zurich, Switzerland, September 6-12, 2014, Proceedings, Part V 13}, pages 740--755. Springer, 2014.

\bibitem[Lipman et~al.(2022)Lipman, Chen, Ben-Hamu, Nickel, and Le]{lipman2022flow}
Yaron Lipman, Ricky~TQ Chen, Heli Ben-Hamu, Maximilian Nickel, and Matt Le.
\newblock Flow matching for generative modeling.
\newblock \emph{arXiv preprint arXiv:2210.02747}, 2022.

\bibitem[Liu et~al.(2023)Liu, Chen, Yuan, Mei, Liu, Mandic, Wang, and Plumbley]{liu2023audioldm}
Haohe Liu, Zehua Chen, Yi Yuan, Xinhao Mei, Xubo Liu, Danilo Mandic, Wenwu Wang, and Mark~D Plumbley.
\newblock Audioldm: Text-to-audio generation with latent diffusion models.
\newblock \emph{arXiv preprint arXiv:2301.12503}, 2023.

\bibitem[Liu et~al.(2024)Liu, Yuan, Liu, Mei, Kong, Tian, Wang, Wang, Wang, and Plumbley]{liu2024audioldm}
Haohe Liu, Yi Yuan, Xubo Liu, Xinhao Mei, Qiuqiang Kong, Qiao Tian, Yuping Wang, Wenwu Wang, Yuxuan Wang, and Mark~D Plumbley.
\newblock Audioldm 2: Learning holistic audio generation with self-supervised pretraining.
\newblock \emph{IEEE/ACM Transactions on Audio, Speech, and Language Processing}, 2024.

\bibitem[Liu et~al.(2022)Liu, Gong, and Liu]{liu2022flow}
Xingchao Liu, Chengyue Gong, and Qiang Liu.
\newblock Flow straight and fast: Learning to generate and transfer data with rectified flow.
\newblock \emph{arXiv preprint arXiv:2209.03003}, 2022.

\bibitem[Lou et~al.(2024)Lou, Meng, and Ermon]{loudiscrete}
Aaron Lou, Chenlin Meng, and Stefano Ermon.
\newblock Discrete diffusion modeling by estimating the ratios of the data distribution.
\newblock In \emph{Forty-first International Conference on Machine Learning}, 2024.

\bibitem[Lu et~al.(2024)Lu, Clark, Lee, Zhang, Khosla, Marten, Hoiem, and Kembhavi]{lu2024unified}
Jiasen Lu, Christopher Clark, Sangho Lee, Zichen Zhang, Savya Khosla, Ryan Marten, Derek Hoiem, and Aniruddha Kembhavi.
\newblock Unified-io 2: Scaling autoregressive multimodal models with vision language audio and action.
\newblock In \emph{Proceedings of the IEEE/CVF Conference on Computer Vision and Pattern Recognition}, pages 26439--26455, 2024.

\bibitem[Mei et~al.(2024)Mei, Meng, Liu, Kong, Ko, Zhao, Plumbley, Zou, and Wang]{mei2024wavcaps}
Xinhao Mei, Chutong Meng, Haohe Liu, Qiuqiang Kong, Tom Ko, Chengqi Zhao, Mark~D Plumbley, Yuexian Zou, and Wenwu Wang.
\newblock Wavcaps: A chatgpt-assisted weakly-labelled audio captioning dataset for audio-language multimodal research.
\newblock \emph{IEEE/ACM Transactions on Audio, Speech, and Language Processing}, 2024.

\bibitem[{MidJourney AI}(2024)]{midjourney2024}
{MidJourney AI}.
\newblock Image generated using midjourney ai, 2024.
\newblock Accessed on November 21, 2024. URL: https://www.midjourney.com/.

\bibitem[Nichol and Dhariwal(2021)]{nichol2021improved}
Alexander~Quinn Nichol and Prafulla Dhariwal.
\newblock Improved denoising diffusion probabilistic models.
\newblock In \emph{International conference on machine learning}, pages 8162--8171. PMLR, 2021.

\bibitem[OpenAI(2023)]{openai2023dalle3}
OpenAI.
\newblock Dall-e 3, 2023.

\bibitem[Peebles and Xie(2023)]{peebles2023scalable}
William Peebles and Saining Xie.
\newblock Scalable diffusion models with transformers.
\newblock In \emph{Proceedings of the IEEE/CVF International Conference on Computer Vision}, pages 4195--4205, 2023.

\bibitem[Podell et~al.(2023)Podell, English, Lacey, Blattmann, Dockhorn, M{\"u}ller, Penna, and Rombach]{podell2023sdxl}
Dustin Podell, Zion English, Kyle Lacey, Andreas Blattmann, Tim Dockhorn, Jonas M{\"u}ller, Joe Penna, and Robin Rombach.
\newblock Sdxl: Improving latent diffusion models for high-resolution image synthesis.
\newblock \emph{arXiv preprint arXiv:2307.01952}, 2023.

\bibitem[Radford et~al.(2021)Radford, Kim, Hallacy, Ramesh, Goh, Agarwal, Sastry, Askell, Mishkin, Clark, et~al.]{radford2021learning}
Alec Radford, Jong~Wook Kim, Chris Hallacy, Aditya Ramesh, Gabriel Goh, Sandhini Agarwal, Girish Sastry, Amanda Askell, Pamela Mishkin, Jack Clark, et~al.
\newblock Learning transferable visual models from natural language supervision.
\newblock In \emph{International conference on machine learning}, pages 8748--8763. PMLR, 2021.

\bibitem[Rombach et~al.(2022)Rombach, Blattmann, Lorenz, Esser, and Ommer]{rombach2022high}
Robin Rombach, Andreas Blattmann, Dominik Lorenz, Patrick Esser, and Bj{\"o}rn Ommer.
\newblock High-resolution image synthesis with latent diffusion models.
\newblock In \emph{Proceedings of the IEEE/CVF conference on computer vision and pattern recognition}, pages 10684--10695, 2022.

\bibitem[Sahoo et~al.(2024)Sahoo, Arriola, Gokaslan, Marroquin, Rush, Schiff, Chiu, and Kuleshov]{sahoo2024simple}
Subham~Sekhar Sahoo, Marianne Arriola, Aaron Gokaslan, Edgar~Mariano Marroquin, Alexander~M Rush, Yair Schiff, Justin~T Chiu, and Volodymyr Kuleshov.
\newblock Simple and effective masked diffusion language models.
\newblock In \emph{The Thirty-eighth Annual Conference on Neural Information Processing Systems}, 2024.

\bibitem[Tang et~al.(2024)Tang, Yang, Zhu, Zeng, and Bansal]{codi}
Zineng Tang, Ziyi Yang, Chenguang Zhu, Michael Zeng, and Mohit Bansal.
\newblock Any-to-any generation via composable diffusion.
\newblock \emph{Advances in Neural Information Processing Systems}, 36, 2024.

\bibitem[Team(2024)]{team2024chameleon}
Chameleon Team.
\newblock Chameleon: Mixed-modal early-fusion foundation models.
\newblock \emph{arXiv preprint arXiv:2405.09818}, 2024.

\bibitem[Tong et~al.(2023)Tong, Malkin, Huguet, Zhang, Rector-Brooks, Fatras, Wolf, and Bengio]{tong2023conditional}
Alexander Tong, Nikolay Malkin, Guillaume Huguet, Yanlei Zhang, Jarrid Rector-Brooks, Kilian Fatras, Guy Wolf, and Yoshua Bengio.
\newblock Conditional flow matching: Simulation-free dynamic optimal transport.
\newblock \emph{arXiv preprint arXiv:2302.00482}, 2\penalty0 (3), 2023.

\bibitem[Vedantam et~al.(2015)Vedantam, Lawrence~Zitnick, and Parikh]{vedantam2015cider}
Ramakrishna Vedantam, C Lawrence~Zitnick, and Devi Parikh.
\newblock Cider: Consensus-based image description evaluation.
\newblock In \emph{Proceedings of the IEEE conference on computer vision and pattern recognition}, pages 4566--4575, 2015.

\bibitem[Yu et~al.(2021)Yu, Li, Koh, Zhang, Pang, Qin, Ku, Xu, Baldridge, and Wu]{yu2021vector}
Jiahui Yu, Xin Li, Jing~Yu Koh, Han Zhang, Ruoming Pang, James Qin, Alexander Ku, Yuanzhong Xu, Jason Baldridge, and Yonghui Wu.
\newblock Vector-quantized image modeling with improved vqgan.
\newblock \emph{arXiv preprint arXiv:2110.04627}, 2021.

\bibitem[Zhang et~al.(2024)Zhang, Zeng, Wang, and Lu]{zhang2024tinyllama}
Peiyuan Zhang, Guangtao Zeng, Tianduo Wang, and Wei Lu.
\newblock Tinyllama: An open-source small language model.
\newblock \emph{arXiv preprint arXiv:2401.02385}, 2024.

\bibitem[Zhou et~al.(2024)Zhou, Yu, Babu, Tirumala, Yasunaga, Shamis, Kahn, Ma, Zettlemoyer, and Levy]{tranfusion}
Chunting Zhou, Lili Yu, Arun Babu, Kushal Tirumala, Michihiro Yasunaga, Leonid Shamis, Jacob Kahn, Xuezhe Ma, Luke Zettlemoyer, and Omer Levy.
\newblock Transfusion: Predict the next token and diffuse images with one multi-modal model.
\newblock \emph{arXiv preprint arXiv:2408.11039}, 2024.

\bibitem[Zhu et~al.(2023)Zhu, Lin, Ning, Yan, Cui, HongFa, Pang, Jiang, Zhang, Li, et~al.]{zhulanguagebind}
Bin Zhu, Bin Lin, Munan Ning, Yang Yan, Jiaxi Cui, WANG HongFa, Yatian Pang, Wenhao Jiang, Junwu Zhang, Zongwei Li, et~al.
\newblock Languagebind: Extending video-language pretraining to n-modality by language-based semantic alignment.
\newblock In \emph{The Twelfth International Conference on Learning Representations}, 2023.

\end{thebibliography}
}

\newpage
\maketitlesupplementary

\appendix



\section{Implementation Details}
\subsection{Dataset}
\label{sec:appendix-dataset}
\begin{table}[]
    \centering
    \begin{tabular}{c|ccc}
         &  Size & Modality  \\
              \hline
      LAION-Aesthetics-3M & 2M* & T,I \\  
   CC12M & 12M & T,I\\  
    COYO-700M(Subset) & 5M  & T,I \\  
     LAION-COCO & 7M & T,I \\  
     SoundNet & 2M & T,A,I$\dag$ \\ 
     VGGSound & 0.2M & T,A,I$\dag$ \\ 
     T2I-2M & 2M & T,I \\ 
    AudioSet & 2M & T,A \\ 
     AudioCaps & 46K & T,A \\ 
     WavCaps & 0.4M & T,A \\ 
     \hline
    \end{tabular}
    \caption{\textbf{List of all datasets used in training. }*Some image URLs are no longer accessible. $\dag$ We generate synthetic captions using BLIP.   }
    \label{tab:all_dataset}
\end{table}

In \cref{tab:all_dataset}, we list the size of all datasets used in training. We filter out all images whose shortest side is less than 256 pixels. To obtain data with all modalities (image, audio, text), we use BLIP to generate synthetic captions for images in the SoundNet\cite{aytar2016soundnet}  and VGGSound\cite{chen2020vggsound} dataset, which are extracted from videos. Since AudioSet only comes with class labels, we use synthetic captions generated by audio-language models provided by AudioSetCaps\cite{bai2024audiosetcaps}.

\subsection{Schedules}
\begin{figure}
    \centering
    \includegraphics[width=1.0\linewidth]{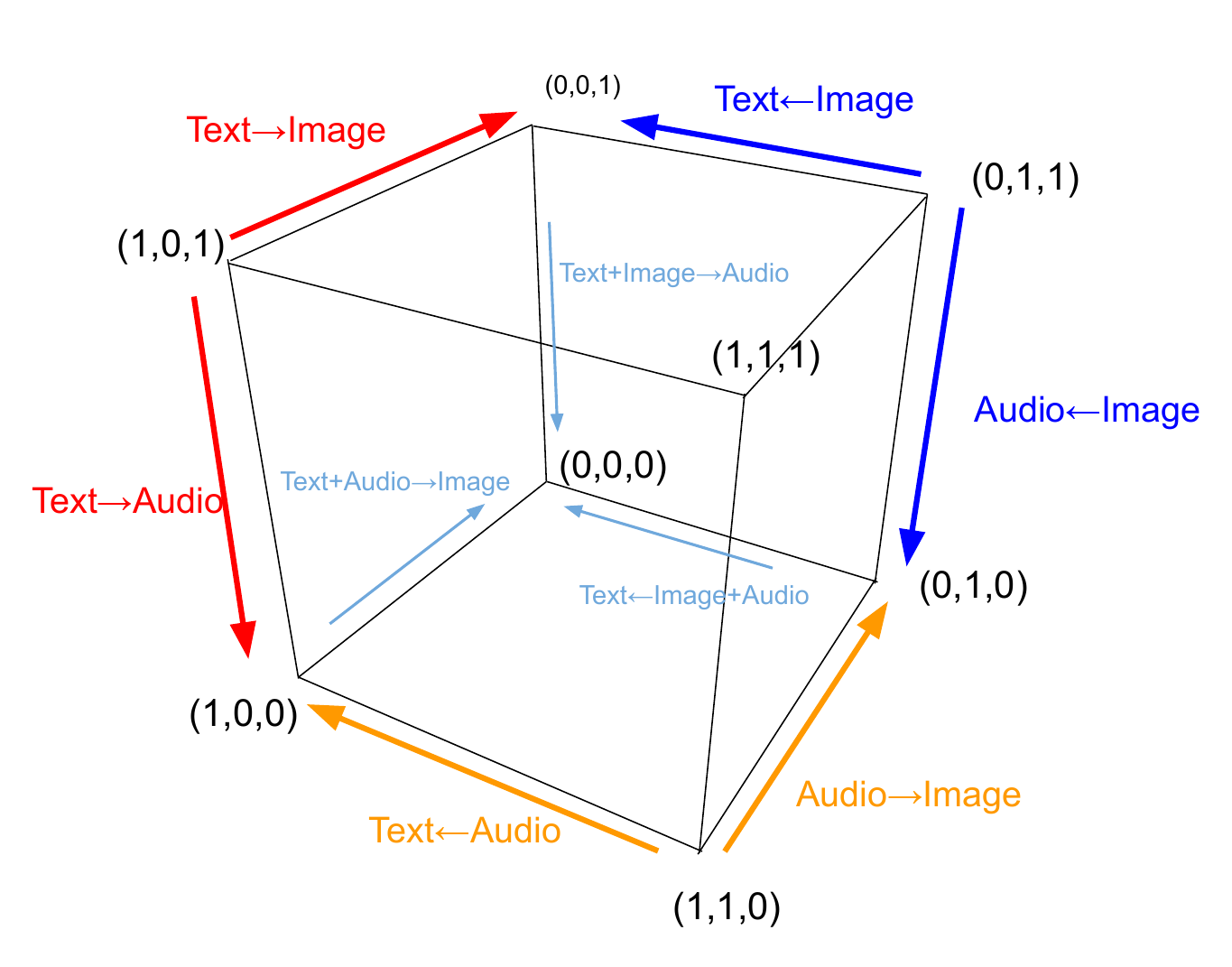}
    \caption{\textbf{Paths encoding different any-to-any generation tasks.} $(t_1,t_2,t_3)$ represents the ``noise level" of image, text and audio modalities. $(0,0,0)$ represents clean (image, text, audio) triplets, and $(1,1,1)$ represents pure Gaussian noise.}
    \label{fig:mm-path-cube}
\end{figure}

Recall from Section \ref{sec:method} that we can represent different tasks with different paths in $[0,1]^3$. We visualize this in \cref{fig:mm-path-cube}. We adopted simple linear tasks for any-to-any generation tasks so that for simple cases like text-to-image and text-to-audio, our formulation matches the standard rectified flow.

\subsection{Training Pipeline}

\begin{figure}
    \centering
    \includegraphics[width=1.0\linewidth]{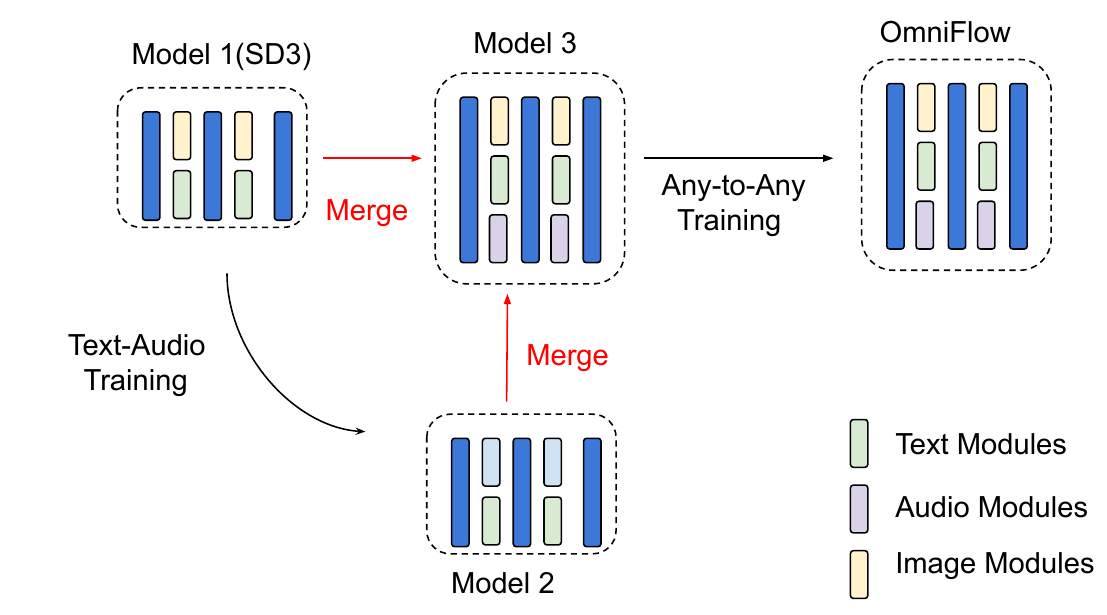}
    \caption{\textbf{Training Pipeline of \ours.} We initialize our model with SD3 (Model 1). We then train the model on text-audio pairs to obtain Model 2. We merge Model 1 and Model 2 to obtain Model 3. The final model is obtained by further training Model 3 on any-to-any generation tasks.}
    \label{fig:training-pipeline}
\end{figure}

We initialize our model with SD3 (Model 1 in \cref{fig:training-pipeline}). We first train the model on text-audio pairs to obtain Model 2. The text branch of Model 2 is initialized with weights of SD3, while the audio branch is randomly initialized. After the training, we merge Model 1, which contains a text branch and an image branch, and Model 2, which contains a text branch and an audio branch, to Model 3, which contains text, image, and audio branches. The text branch of Model 3 is obtained by averaging the weights of the text branches from Model 1 and 2. Finally, we train the Model 3 on all datasets mentioned in \cref{sec:appendix-dataset}. This training pipeline is illustrated in \cref{fig:training-pipeline}.

We train Model 2 for 100k steps and Model 3 for 150k steps. We use 8 A6000 GPUs with a per GPU batch size of 8. We use AdamW optimizer with a learning rate of 1e-5 for Model 2 and 5e-6 for Model 3. The learning rate undergoes a linear warmup in the first 1000 steps and a cosine decay throughout the rest of the training. We adopt exponential moving average (EMA), which are updated every 100 training steps with a decay factor of 0.999. 

\begin{figure}
    \centering
    \includegraphics[width=1.0\linewidth]{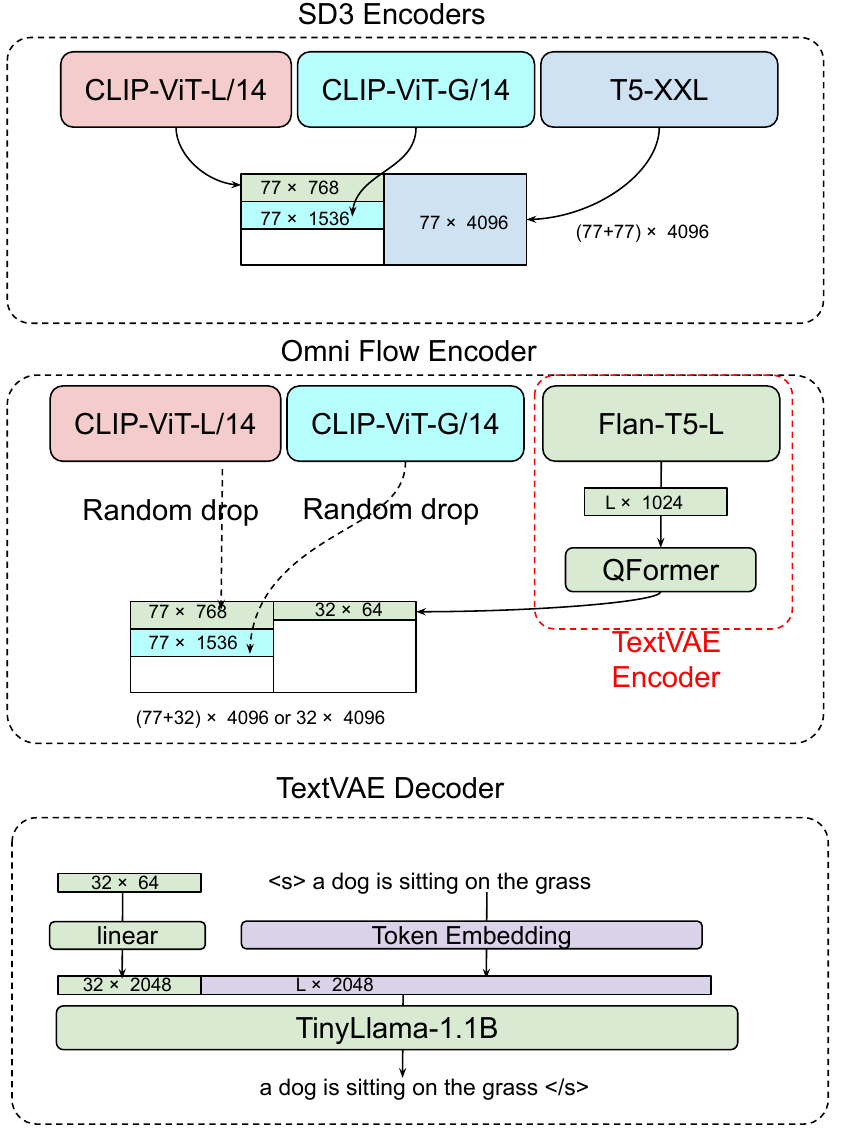}
    \caption{\textbf{Architecture of Text VAE and Text Encoders in \ours.} SD3 (Top) uses three text encoders: CLIP-L, CLIP-G, and T5-XXL. \ours~(Middile) replaces the 4.7B T5-XXL with a VAE encoder based on Flan-T5-L. CLIP encoders become optional and are not used for tasks without clean text inputs. The decoder of VAE (Bottom) is based on TinyLlama-1.1B. The VAE embedding is used as the prefix for decoding. }
    \label{fig:textvae}
\end{figure}

\subsection{Text VAE}

We train a text VAE on caption data using Flan-T5-L \cite{chung2024scaling}. Recall that SD3\cite{esser2024scaling} makes use of three text encoders: CLIP-L, CLIP-G and T5-XXL. We replace the 4.7B T5-XXL with Flan-T5-L \cite{li2020optimus} to save computation cost and use it as part of a text VAE. Specifically, given an input caption of length $L$, it is first encoded by Flan-T5-L to obtain a vector of size $L\times1024$. We then pass it to a QFormer\cite{li2023blip} and obtain an output vector of size  $32\times64$. This vector is used as the VAE embedding. In the decoding process, the VAE embedding is first processed by a linear projection layer to obtain a vector of size $32\times2048$. This is used as the prefix embedding for a TinyLlama-1.1B decoder \cite{zhang2024tinyllama}. These architecture designs are shown in \cref{fig:textvae}. Note that while we introduced a 1.1B text-decoder, the overall system actually has fewer parameters since we replaced the 4.7B T5-XXL with a 783M Flan-T5-L.

We employ the auto-encoding training objective of OPTIMUS \cite{li2020optimus}. We freeze the Flan-T5-L encoder and fine-tune the QFormer and TinyLlama decoder end-to-end. We train the text VAE on all caption data mentioned in \cref{sec:appendix-dataset} for 2 epochs, with a learning rate of 1e-5, a global batch size of 256 using AdamW optimizer.

When using the VAE encoder as the text encoder of \ours, we pad the embedding to 4096 with zeros to maintain the input dimension of SD3. Additionally, we also incorporate the CLIP-L and CLIP-G encoders of SD3 as auxiliary text encoders to stabilize the training. We apply random dropout to these encoders during the training. During the inference, the CLIP encoders are not used if the input does not contain clean texts (e.g. Image-to-Text task).

\subsection{Audio VAE}

We directly adapt the audio VAE used by AudioLDM \cite{liu2023audioldm}. In particular,we adopt the same vocoder and preprocessing pipeline as AudioLDM2. We use HiFiGen as VAE, which is used in AudioLDM and AudioLDM2. We use AudioLDM2's checkpoint. We also explored AudioMAE, but found it to perform significantly worse as measured by FAD (2.03 vs 1.79). 

\subsection{Omni-Transformer}

We followed the architectural design of SD3 for image and text modules and initialize them with SD3 weights. The audio modules are initialized with identical setup to the image modules. Specifically, it has 24 layers and a hidden size of 1536. The positional embedding layer has a patch size of 2. Since the audio VAE outputs a feature map of dimension $256\times16$, the positional embedding layer will convert each audio to a sequence of length $128\times8=1024$. 

\subsection{Pooled Conditional Embeddings}

SD3 makes use of additional pooled embeddings from CLIP-ViT-L/14 and CLIP-ViT-G/14 in addition to the sequence embeddings. We maintain them as is, with additional dropout during the training. We additionally incorporate an Audio Encoder to create pooled embeddings for audio inputs \cite{zhulanguagebind}. These embeddings are not used when clean data of respective modality is not available.

\subsection{Baselines}

In this section, we describe the specific variants studied in \cref{tab:ablation}. Except for the discrete text diffusions (SEDD and MDLM), these variants fit into the unified formulation of \cref{eq:unified-repr} by varying its parameters.

\textbf{linear} is a variant of DDPM used in LDM \cite{rombach2022high}.  It discretizes the timesteps to $0,1...T-1$ and uses the formulation $b_t=\sqrt{1-\alpha_t^2}$, where $a_t=\sqrt{\prod_{i=0}^t(1-\beta_i)}$, and $\beta_t=(\sqrt{\beta_0}+\frac{t}{T-1}(\sqrt{\beta_{T-1}}-\sqrt{\beta_0}))^2$. We explored $\epsilon$-prediction and $v$-prediction objectives for this variant.

\textbf{cosine} is defined by the forward process 

\begin{equation}
    x^t=\cos(\frac{\pi}{2}t)x^0+=\sin(\frac{\pi}{2}t)x^1
\end{equation}

The weighting function is $w_t=e^{-\lambda_t/2}$ for $v$-prediction objectives\cite{ho2022imagen}.

\begin{figure}
    \centering
    \includegraphics[width=1.0\linewidth]{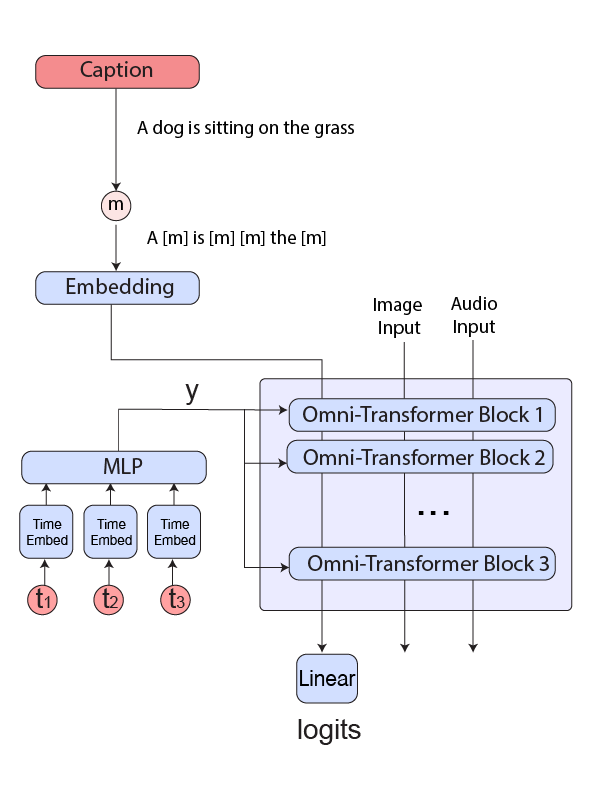}
    \caption{\textbf{Discrete Diffusion Variant of \ours.} In this setup, we remove the text VAE and directly pass token embedding to the Omni-Transformer layers. ``[m]" indicates a mask token.}
    \label{fig:text-diffusion}
\end{figure}

\textbf{SEDD and MDLM} are recently proposed discrete text-diffusion models. We consider MDLM\cite{sahoo2024simple} and the absorbing state variants of SEDD\cite{loudiscrete} in our experiments.\footnote{SEDD also has a uniform variant, where the tokens are not replaced with a "[MASK]" token, but a randomly token sampled from the vocabulary.} These models directly define a forward process in the discrete token space, where clean text tokens are progressively replaced with a special ``[MASK]" token. We adapt our implementation for these methods by removing the text VAE and introducing a token embedding layer. This design is shown in \cref{fig:text-diffusion}.

\section{Additional Discussions}

\subsection{Sampling}

\ours~does not directly model the marginals of two modalities. For example, given three modalities $(x_1,x_2,x_3)$, it does not directly model $p(x_1^0|x_2^0)=\int_{x_3^1\in\mathcal{\mathbb{R}}^{d_3}} p(x_1^0,x_3^1|x_2^0)  dA $,  where $d_3$ is the dimension of $x_3^1$. Integrating over $x_3^1$ is infeasible. Instead, we sample $p(x_1^0,x_3^1|x_2^0)$ by first sample $x_3^1\sim q(x_3^1|x_2^0)=\mathcal{N}(0,I)$ and sample $p(x_1^0|x_3^1,x_2^0)$ using path from (1,0,1) to (0,0,1). 

\subsection{Necessity of text, image, audio triplets. }

\begin{figure}[t]
    \centering
    \includegraphics[width=1.0\linewidth]{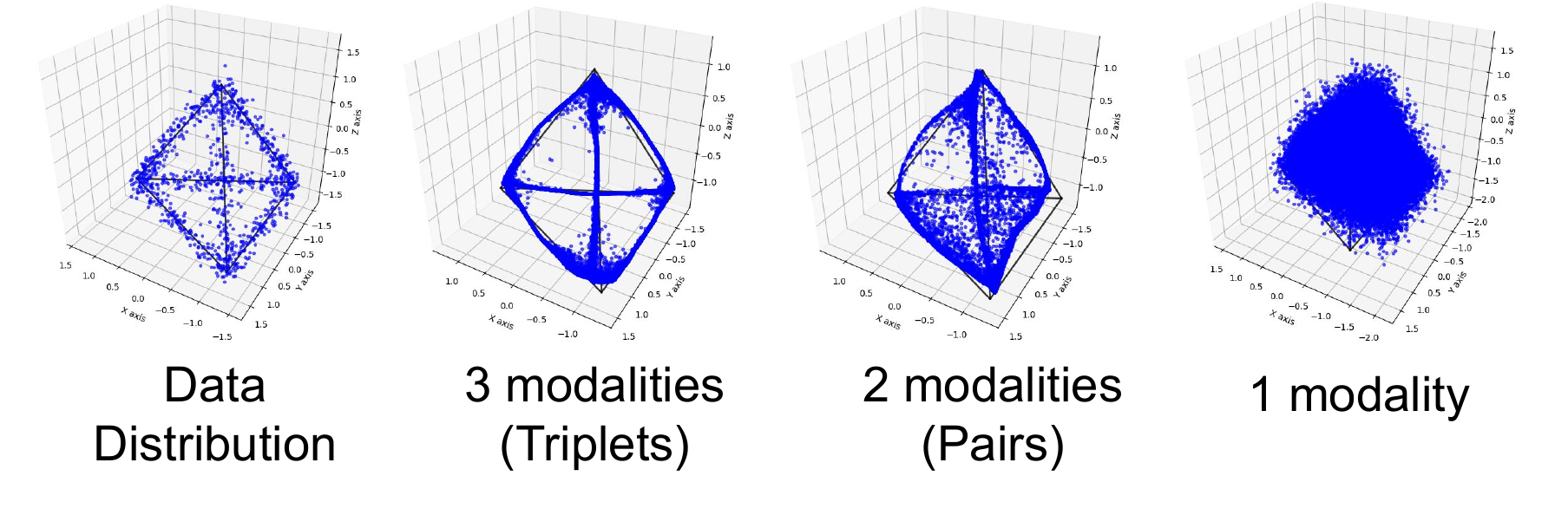}
    \caption{\textbf{Synthetic Experiments on three 1D-modalities.} We consider the joint distribution of three toy modalities ($x_1,x_2,x_3$), each represented by a vector of dimension 1. Hence, a triplet consisting of three modalities be represented by a point in $\mathbb{R}^3$ We assume the joint distribution is a uniform distribution in the neighborhood of tetrahedron (Left). We experiment with training \ours using triplets, pairs, and only individual modalities. Models trained with triplets of three modalities best represent the original distribution.  }
    \label{fig:distributionl}
\end{figure}

Compared with previous works such as CoDi\cite{codi} which uses weighted average of embeddings to mix multiple input modalities, \ours~requires directly training on triplets consisting of all modalities (image, text, audio). To study the necessity of this requirement, we conduct synthetic toy examples on three modalities $(x_1,x_2,x_3)$, each represented by a one-dimensional vector.  A triplet of three modalities can then be represented by a point $(X,Y,Z)$ in 3D space. We show this experiments in \cref{fig:distributionl}. We assume the ground truth data distribution follows a uniform distribution in a small neighborhood adjacent to a tetrahedron (Leftmost Figure).  We experiment with training an 8-layer MLP with triplets $(x_1,x_2,x_3)$ (Second-Left Figure), pairs of $(x_1,x_2)$,$(x_1,x_3)$,$(x_2,x_3)$ (Second-Right Figure), and only individual modalities $(x_1),(x_2),(x_3)$ (Rightmost Figure). For each model, we plot 50k samples generated by the model. Qualitatively, models trained on triplets best represent the data distribution. This makes sense as pairs are essentially projections on XY, XZ, YZ planes and individual modalities are projections on X, Y, Z axis. These projections are not sufficient to recover the original distribution represented in this 3D space.  

\section{Quantative Text Evaluation}

\begin{table*}[ht!]
\centering
\begin{tabular}{lcc|cc|cc}
& & & \multicolumn{2}{c|}{\textbf{AudioCaps}}& \multicolumn{2}{c}{\textbf{COCO-Karpathy}}
\\
 & \textbf{Images} & \textbf{Parms}. & \textbf{CLAP$\uparrow$} & \textbf{CIDEr$\uparrow$} & \textbf{CLIP$\uparrow$} & \textbf{CIDEr$\uparrow$} \\ \hline
  \multicolumn{7}{c}{\textit{Specialist}} \\
  BLIP-2\cite{li2023blip} & 129M & 2.7B & - & - & - & 145.8 $\ddagger$ \\
  SLAM-AAC\cite{chen2024slam} & - & 7B & - & 84.1$\ddagger$ & - & - \\
  \hline
 \multicolumn{7}{c}{\textit{Generalist}} \\
\ours & 30M & 3.4B & \textbf{0.254} & \textbf{48.0} & 26.8 & \textbf{47.3} \\ 
CoDi $\dag$ & 400M & 4.3B & 0.206 & 7.9 & 25.9 & 17.2 \\ 
Unidiffuser $\dag$ & 2B & 0.9B & - & - & \textbf{29.3} & 20.5 \\ \hline
\textcolor{gray}{UIO2-XXL} & \textcolor{gray}{1B*} & \textcolor{gray}{6.8B} & \textcolor{gray}{-} & \textcolor{gray}{48.9} & \textcolor{gray}{-} & \textcolor{gray}{125.4*}  \\ 
\textcolor{gray}{Transfusion} & \textcolor{gray}{3.5B} & \textcolor{gray}{7B} & \textcolor{gray}{-} & \textcolor{gray}{-}& \textcolor{gray}{-} & \textcolor{gray}{35.2}  \\ 
\hline
\end{tabular}
\caption{\textbf{X-to-Text Performance comparison on AudioCaps and COCO Captions.} * UIO2's training data includes COCO. The fine-tuning dataset also includes 53M image understanding data, including 14 image captioning datasets. $\dag$ evaluated with official checkpoints. $\ddagger$ fine-tuned on respective datasets (COCO and Audiocaps). }
\label{tab:caption-performance}
\end{table*}

We report quantitative results of image captioning on COCO-Karpathy-Test dataset and audio captioning on Audiocaps dataset. We report CLIP score, CLAP score, and CIDEr\cite{vedantam2015cider} on these two benchmarks. We compare against generalist models such as CoDi and Uni-Diffuser. Uni-Diffuser, released two checkpoints v0 and v1, where v1 is fine-tuned on internal data. We compare against v0 for a fairness. \ours~outperforms CoDi on both tasks, and outperforms UniDiffuser in CIDEr score (+26.8). It has a lower CLIP score (-2.5). We consider the performance of \ours~as competitive, considering \ours~is trained on significantly less data than UniDiffuser and can also perform audio captioning task.  We note that the performance of generalist models significantly lags behind specialist models that are fine-tuned of respective datasets, suggesting rooms for further improvements. We provide further discussion in the limitation section.

\section{Benefits of Multi-Task Training}

In this section, we discuss if joint training on multiple modalities benefit single-task performance. We provide additional results in \cref{tab:results_joint} by comparing a model trained on all tasks with a model only trained on a subset of tasks. We show that Omniflow was able to leverage the training data of related tasks (e.g. T2A, I2A) and boost individual performance. In additional to results presented in Tab.R1, we also observe improvements in image generation, where OmniFlow generate high fidelity A2I outputs even though A2I datasets consist of low-res videos (+1.22 Aesthetic score), thanks to high-fidelity T2I data.

\begin{table}[t]
    \centering
    \begin{tabular}{HHc|c H}
       Model & IR. $\uparrow$ & \textbf{Training Data} &\textbf{FAD $\downarrow$ }& A2I $\uparrow$  \\
    \hline
        OmniFlow & \textbf{1.20} & OmniFlow  & \textbf{1.83} &\textbf{x} \\
        \hline
         SDv1.5 & 0.33 & I2A,A2I,T2A,A2T& 1.89 & 0.52 \\
        SDXL  &  0.93 & I2A,A2I  & 2.03  & 0.48 \\
        SD3  &   1.09 & I2A-Only  & 2.05 & -\\

        \hline
         
    \end{tabular}
    \caption{ \textbf{Performance of Various Training Data Compositions.} We compare FAD scores for Image to Audio (I2A) under different setup on VGGSound.   }
    \label{tab:results_joint}
\end{table}

\section{Additional Qualitative Results}

\subsection{Text-to-Image}
\begin{figure*}[h!]
    \centering
    \includegraphics[width=0.9\linewidth]{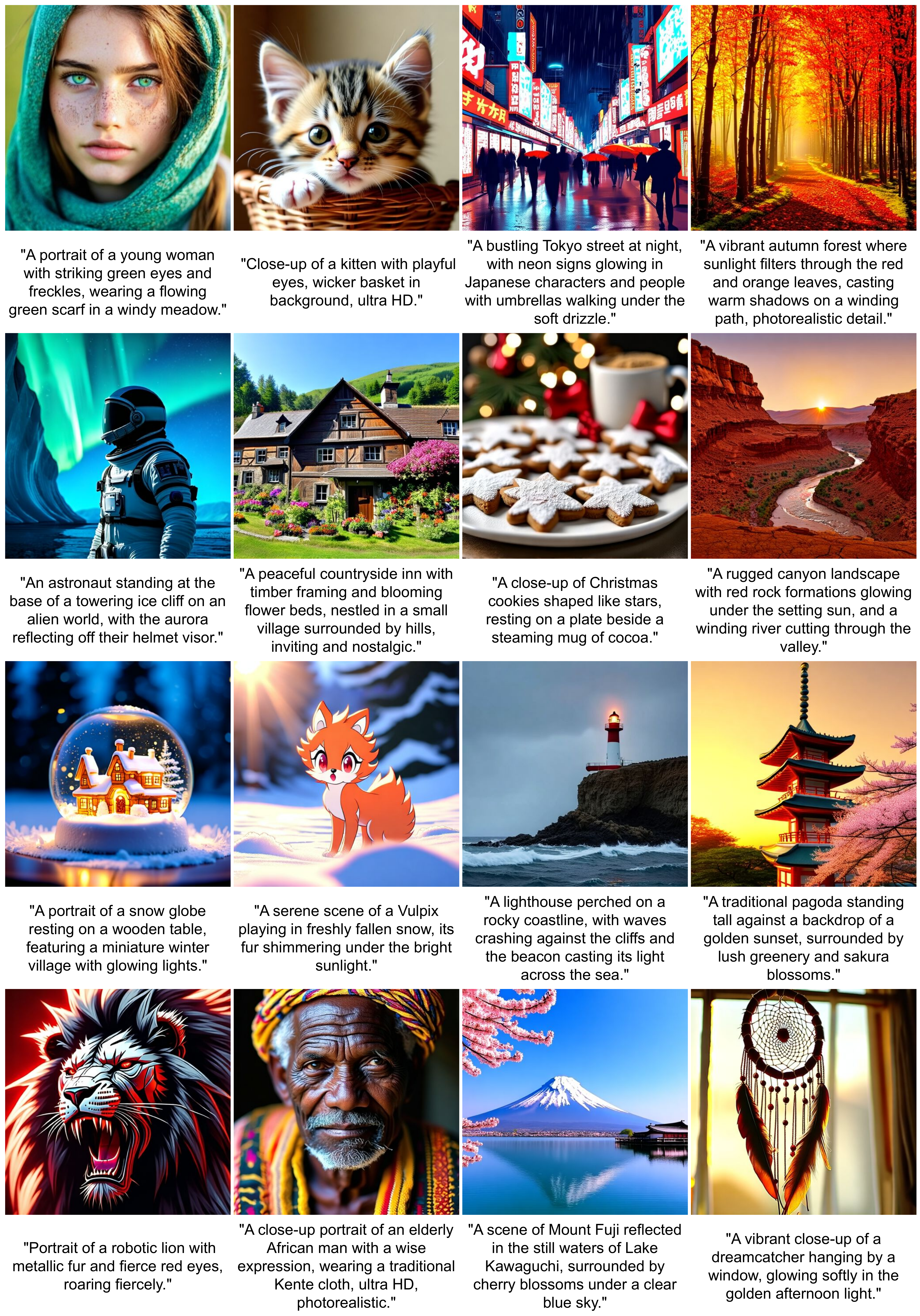}
    \caption{\textbf{Qualitative examples of the text-to-image capability of \ours.}}
    \label{fig:sup_t2i_1}
\end{figure*}

\cref{fig:sup_t2i_1} demonstrates a range of qualitative text-to-image examples for \ours. We depict a wide variety of people, scenes and objects to demonstrate the robustness of our approach.

\subsection{Image-to-Text}
\begin{figure}[h]
    \centering
    \includegraphics[width=1.0\linewidth]{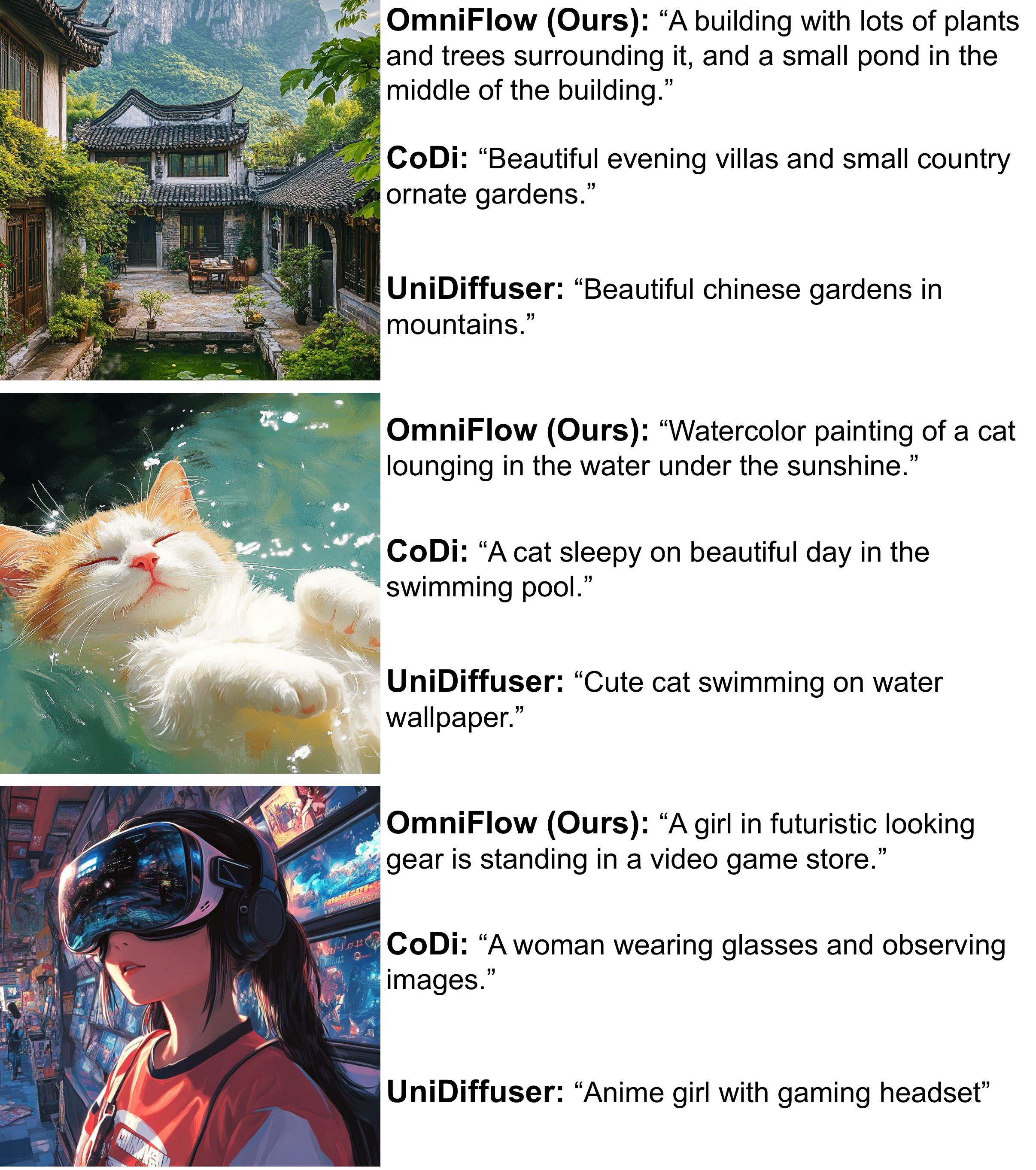}
    \caption{\textbf{Qualitative comparison of \ours~with baselines on image-to-text generation.} Images are provided from the Midjourney Explore page \cite{midjourney2024}.}
    \label{fig:side-by-side-i2t}
\end{figure}

We provide a side-by-side image-to-text comparison between \ours~, CoDi \cite{codi} and UniDiffuser \cite{unidiffuser} using synthetic high quality images from the Midjourney Explore page \cite{midjourney2024}in \cref{fig:side-by-side-i2t}.

\subsection{Audio-to-Text}
In \cref{tab:audiocaps-a2t}, we show qualitative results on Audiocaps audio-to-text task. \ours~can generation captions that match the ground truth. While CoDi can correctly grasp the main objects in the audio such as ``car", ``bird", ``sheep", ``computer", it struggles with generating captions that accurately reflect the scene.

\subsection{Text-VAE AutoEncoding}

In \cref{tab:text_vae_reconst}, we show reconstruction examples of Text VAE. The reconstruction mostly adheres to the semantics of the ground truth, with minor differences. For example, it may change ``well-furnished" to ``well-decorated".

\begin{table*}[ht!]
\centering
\begin{tabular}{|c|p{4.5cm}|p{4.5cm}|p{4.5cm}|}
\hline
\textbf{ID} & \textbf{CoDi} & \textbf{\ours} & \textbf{GT} \\ \hline
yVjivgsU2aA & Four car driver trying forcoming for a speeding car. & A race car engine revs and tires squeal. & An engine running followed by the engine revving and tires screeching. \\ \hline
8F-ndyrEWJ8 & Fire police cars stop and red traffic on different highway. & A fire siren goes off loudly as a man shouts and a low hum of an engine is running throughout the whole time. & A distant police siren, then racing car engine noise, and a man calling in police code over his radio. \\ \hline
350OCezayrk & Four motor car driving for completing an automobile service. & A vehicle engine is revving and idling. & A motor vehicle engine starter grinds, and a mid-size engine starts up and idles smoothly. \\ \hline
LCwSUVuTyvg & Door, a blue hat and winter jacket. & A door is being slammed. & Glass doors slamming and sliding shut. \\ \hline
7XUt6sQS7nM & The sheep of the woman are the sheep of the sheep. & Multiple sheep bleat nearby. & A sheep is bleating and a crowd is murmuring. \\ \hline
PVvi2SDOjVc & Car going for a car coming home. Three cars coming for a blue car coming down a road after the highway. & A car horn beeps. & A car engine idles and then the horn blows. \\ \hline
Z\_smJ66Tb3c & Men in the bird while the man in the boat. & Two men talk over blowing wind and bird chirps. & A man is speaking with bird sounds in the background followed by a whistling sound. \\ \hline
CMNlIW6Lkwc & Two men in the fire and two men are coming towards the other man in the game. & A man speaks, followed by a loud bang and people laughing. & A man talking as a camera muffles followed by a loud explosion then a group of people laughing and talking. \\ \hline
JQz40TkjymY & Writing computers for people in writing. & Typing on a computer keyboard. & Typing on a computer keyboard. \\ \hline
U90e2P9jy30 & A man shouts the word to the person on the sidewalk to walk to get him to the door the hand to fall down on the sidewalk in. & Basketballs being dribbled and people talking. & Several basketballs bouncing and shoes squeaking on a hardwood surface as a man yells in the distance. \\ \hline
5I8lmN8rwDM & Stationary fire drill technician drilling down a hose pipe while wearing safety gear. Railroad safety drill for motorcycle with hose or oil checking equipment. & A drill runs continuously. & Drilling noise loud and continues. \\ \hline
NlKlRKz8OKI & Birds on blue birds. & A woman talks and then an animal chewing. & A woman speaks with flapping wings and chirping birds. \\ \hline
\end{tabular}
\caption{\textbf{Qualitative comparisons of CoDi and \ours~ on Audiocaps audio captioning task.} Audios are randomly sampled. Audiocaps provide five ground truth captions per audio. For better presentation, we only list one in this table.  }
\label{tab:audiocaps-a2t}
\end{table*}

\begin{table*}[ht!]
\centering
\begin{tabular}{|p{0.45\textwidth}|p{0.45\textwidth}|}
\hline
\textbf{Reconstruction} & \textbf{GT} \\ \hline
Crispy chicken tenders alongside a portion of a bbq sauce. & Crispy chicken tenders alongside a portion of bbq sauce. \\ \hline
A \textcolor{red}{well-furnished} living room with a patterned curtain \textcolor{red}{rod}, a small white side table holding a vase of flowers, and a tufted gray sofa. & A \textcolor{red}{well-decorated} living room with a patterned curtain \textcolor{red}{panel hanging from the window}, a small white side table holding a vase of flowers, and a tufted gray sofa. \\ \hline
A young man wearing a black shirt and holding an American flag. & A young man wearing a black shirt and holding an American flag. \\ \hline
An artistic painting of a futuristic city by the water. & An artistic painting of a futuristic city by the water. \\ \hline
Cozy and \textcolor{red}{well-designed} living room with a green velvet sofa, glass coffee table displaying potted plants, and a large skylight overhead. & Cozy and \textcolor{red}{stylish} living room with a green velvet sofa, glass coffee table displaying potted plants, and a large skylight overhead. \\ \hline
A silver \textcolor{red}{Audi Rs4} sedan driving on the passenger side near a mountainous coastline. & A silver \textcolor{red}{Acura RLX} sedan driving on the passenger side near a mountainous coastline. \\ \hline
\end{tabular}
\caption{\textbf{Text VAE reconstruction results. We show reconstruction results (Left) and the ground truth text (Right).}The reconstruction mostly adheres to the semantics of the ground truth, with minor differences.  }
\label{tab:text_vae_reconst}
\end{table*}

\section{Limitations} 

On text generation tasks, our model's performance is not state-of-the-art and has considerable room for improvements. We believe this is the side effect of incorporating large-scale data with many noisy texts of different styles (e.g. alt texts, human written prompts) that differs from the distribution of standard benchmark datasets such as MSCOCO. Additionaly, for image-to-text task specifically, \ours~is exposed to considerably less image-text pairs (30M) during the training compared with previous generalist models such as CoDi(400M) and UniDiffuser(2B). There is also the question of balancing datasets of different caption qualities. For example, WavCaps is a weakly-labeled dataset, but is 10x larger than higher quality AudioCaps. Additional consideration is required in order to generate captions that can achieve high scores on audiocaps benchmark. Despite these limitations, we show that \ours~can generate reasonable image and audio captions through quantitative and qualitative experiments. Our work focuses on develop an effective recipe for any-to-any generalist models. We leave optimizing for text generations to future works. 

On Image generation tasks, while \ours~can generate high quality images, it has the same limitations as any text-to-image models. For example, it may inherit unintended biases from the training dataset. It may also struggle in prompts that the vanilla SD3 model also struggles with.

\begin{figure}[ht!]
    \centering
    \includegraphics[width=1.0\linewidth]{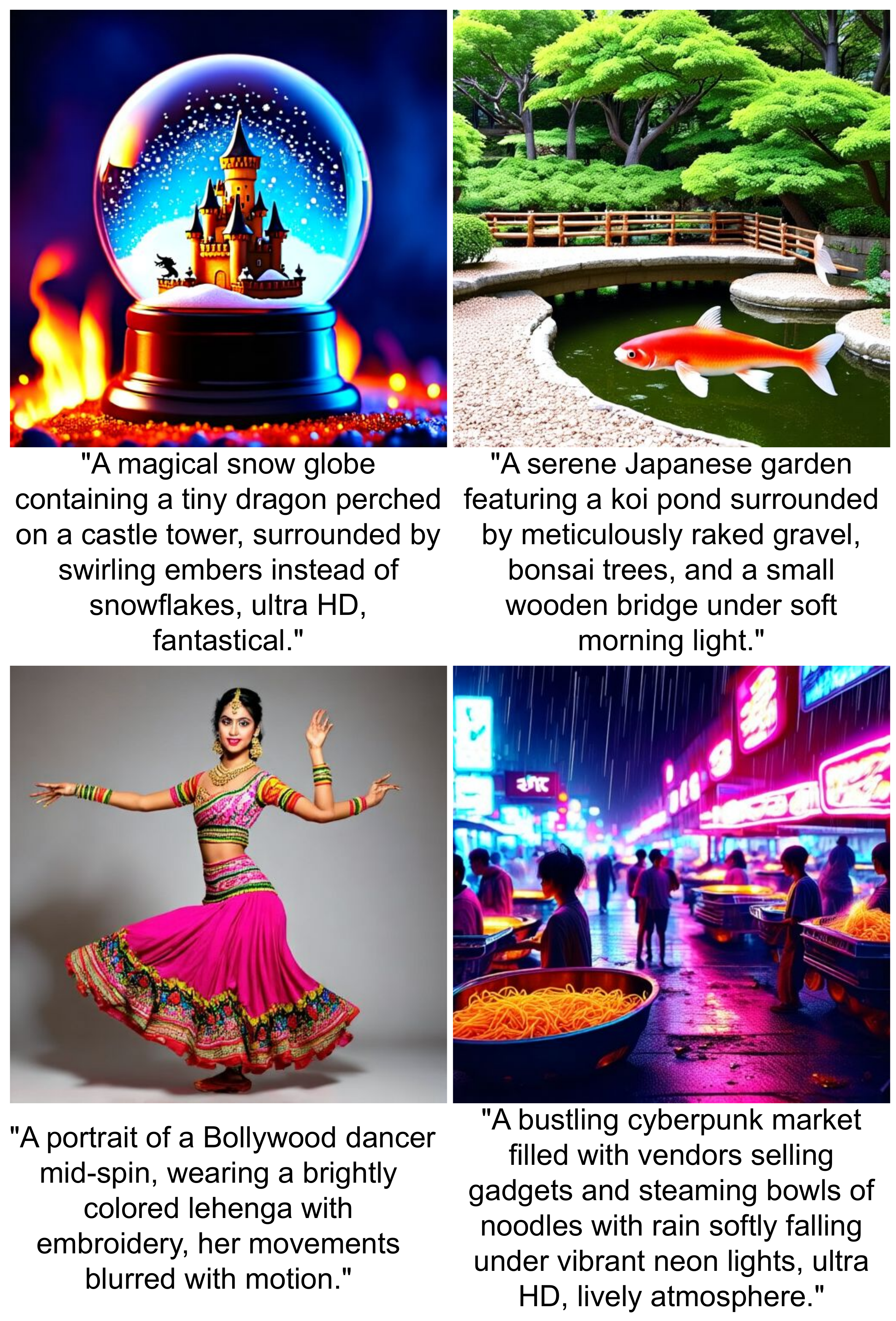}
    \caption{\textbf{Examples of failure cases encountered during the text-to-image generation process of \ours.}}
    \label{fig:failure_cases}
\end{figure}

\section{Miscellaneous }
\subsection{Reproducibility of CoDi}
To accurately reproduce the results of CoDi \cite{codi}, we follow the weights and instructions as indicated in the i-Code-V3 GitHub repository \footnote{https://github.com/microsoft/i-Code/tree/main/i-Code-V3}. However, we encounter reproducibility issues, similar to open issues reported by others, which have remained unresolved \footnote{https://github.com/microsoft/i-Code/issues/134}.

\section{Reproducibility Statement}
All dataset used in this work are public and accessible from the Internet, except for synthetic captions of SoundNet and VGGSound we generated. We have release the code, checkpoints, and generated captions for these two dataset.

\section{Failure Cases}
In \cref{fig:failure_cases} we present several failure cases of \ours~ when performing text-to-image generation. In the snow globe example, the model fails to interpret the prompt specifying "swirling embers instead of snowflakes," mistakenly generating snow instead. Another issue arises with the dancer, where the prompt "movements blurred with motion" is inaccurately represented as an additional arm. Lastly, the Koi pond and ramen examples highlight unnatural outputs, with the former resembling a poorly edited image of a fish in a pond and the latter depicting oversized bowls of noodles placed unnaturally on the street.

\end{document}